\documentclass[5p,review]{elsarticle}

\usepackage{fouriernc}

\usepackage{graphicx}
\usepackage{subfigure}
\usepackage{hyperref}

\usepackage{amsmath}\allowdisplaybreaks
\usepackage{amssymb}
\usepackage{amsthm}
\usepackage{amsfonts}
\usepackage{physics}
\usepackage{mathrsfs}
\usepackage[dvipsnames]{xcolor}
\usepackage{array}
\usepackage{makecell}
\usepackage{comment}
\usepackage{empheq}
\usepackage{verbatim}
\usepackage{multirow}
\usepackage{hhline}
\usepackage[pscoord]{eso-pic}
\usepackage{enumitem}
\setlist[itemize]{noitemsep,nolistsep}
\usepackage{xfrac}
\usepackage{microtype}
\usepackage[normalem]{ulem}

\theoremstyle{remark}

\newtheorem*{remark*}{Remark}
\newcommand{\Rn}[1]{\uppercase\expandafter{\romannumeral #1\relax}}

\newcommand\numberthis{\addtocounter{equation}{1}\tag{\theequation}}

\begin{document}

\begin{frontmatter}
\title{Off the Normal Path: Learning Spatial Density Models of Node Mobility}
\author[mac]{Wanxin Gao\corref{cor1}}
\cortext[cor1]{Corresponding author}
\ead{wanxingao@cityu.edu.mo}
\author[uofa]{Ioanis Nikolaidis}
\ead{nikolaidis@ualberta.ca}
\author[uofa]{Janelle Harms}
\ead{janelleh@ualberta.ca}

\address[mac]{Faculty of Data Science,
City University of Macau,
Macau, China}
\address[uofa]{Department of Computing Science, University of Alberta, Edmonton, Canada}

\begin{abstract}
We consider the problem of learning models of spatial density functions, representing the steady-state density of mobile nodes moving on a two-dimensional terrain. Deriving such models can assist in network design and optimization problems, e.g., by accelerating the computation of the density function during a parameter sweep. We address the question of applicability of off-the-shelf mixture density network models and of, two varieties of, normalizing flows for the description of mobile node density over a disk. We introduce the use of M\"obius distributions to retain symmetric spatial relations. Our results indicate that mixtures of M\"obius distributions provide interpretable, parsimonious models for the studied steady state density distributions, that match or outperform the alternatives.

\end{abstract}

\end{frontmatter}

\section{Introduction}

Over decades, the data communications research community has been well-served by simulation techniques. The main use of simulations is to provide a computational alternative to derive performance metrics when it is prohibitive to build (or compute) an exact analytical model, let alone implement a complete proof-of-concept systems to study its performance. Simulation has also been demonstrated to be useful as a component of larger, usually optimization, strategies. In the latter case, one can think of a simulation as a ``subroutine'' invoked by, e.g., a search \cite{fuHandbookSimulationOptimization2015} or, more recently, a reinforcement learning \cite{gosaviSimulationBasedOptimizationParametric2015,ergunSurveyHowNetwork2023} strategy. Also, entire sub-fields, such as that of ``digital twins'' heavily depends on simulation modules \cite{thelenComprehensiveReviewDigital2022} as the means to form predictions for the actual systems they mirror.

The scenario we consider is that of using simulations as subroutines of a broader, e.g., RL, computational task, and specifically simulations from which we can derive the steady state distributions of wireless, mobile, nodes operating under non-trivial mobility scenarios. %
Our guiding example is a mobility model that includes concepts of points of attraction and limited resources, e.g., energy, of the mobile nodes to require occasional replenishment.

\begin{figure}[t]
 \centering
 \subfigure[]{
 \label{fig:exact:a}
 \includegraphics[width=2.8in]{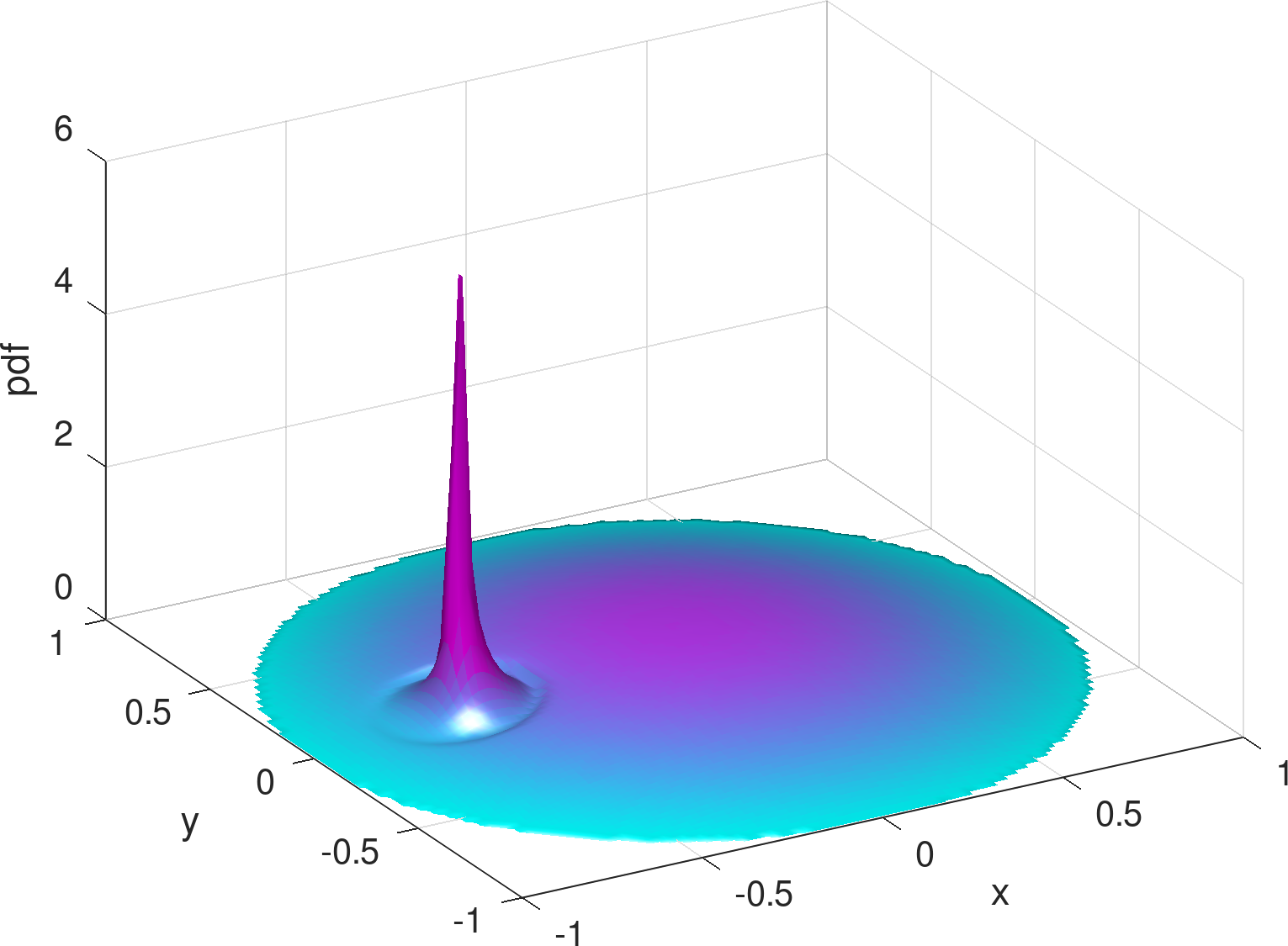}}
 \subfigure[]{
 \label{fig:exact:b}
 \includegraphics[width=2.25in]{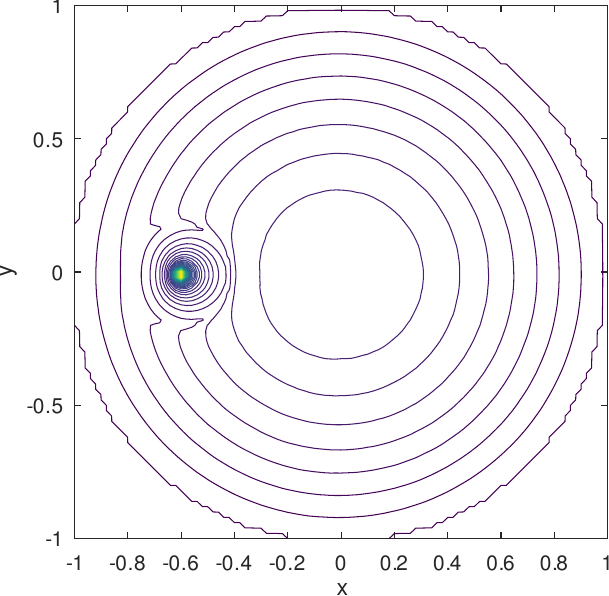}}
 \caption{Example of the exact solution for the node densities approximated. The supporting disk, centered at (0,0), has a radius of 1.}
 \label{fig:exact}
\end{figure}

Existing mobility models range from simple random waypoint (RWP) to fairly elaborate models where nodes are attracted to specific regions/points because of particular spatial features. Closed-form analytical descriptions of the density are rarely available---one noted exception is the RWP density \cite{hyytia2006spatial}. An example of non-trivial density which lacks a closed-form solution and is computationally demanding even for its numerical approximation, is the one derived from the work in \cite{9678073} and shown in the example of Figure \ref{fig:exact}. Zooming into the features of the distribution reveals subtleties like the density ``peak'' drops in a non-symmetric manner to the left and to the right of the peak, while at some distance around this peak there exists a ``dip'' in the density. The notable background ``dome'' is the RWP residual component of the distribution. It is distributions such as the one in Figure \ref{fig:exact} that we wish to model via a learning model. 

 With recent developments in machine learning, a question naturally arising is whether we can forego computationally expensive simulations like those we encounter in user mobility scenarios, with a ``learned'' model of the steady state distribution, i.e., of the result the simulations were providing us. For this approach to be useful as an alternative to simulation, it is implied that the learned models should be built using a \emph{small training set},  yet be powerful in generalization.

To stay consistent with the assumptions made in the literature on the subject of mobility models \cite{leboudecPerfectSimulationStationarity2005}, we consider distributions supported on a two-dimensional disk. A disk is a rough approximation of an enclosed, spatially relevant area, be it a city or a neighborhood, i.e., conceptually surrounding a ``center'' which is a common feature in urban settings, hence its overall appeal in mobility modeling. In summary, we are looking at density models that capture non-trivial shapes of spatial distributions over a disk.

To derive a learning model, we specifically consider two general strategies, ``classical'' mixture density models \cite{bishopMixtureDensityNetworks1994} and a normalizing flow (NF) approach \cite{papamakariosNormalizingFlowsProbabilistic2021}. The paucity of training data can be a concern for the applicability of the more elaborate models, such as NFs. Furthermore, we mainly consider two different kinds of distributions, Gaussian (with and without truncation) as well as families of distributions derived from the beta distribution, and specifically the M\"obius distribution \cite{jonesMobiusDistributionDisc2004}, as it is readily able to capture symmetries present on a disk.

One standard approach that we consider are mixture density networks (MDNs) \cite{bishop1994mixture}, starting by considering typical Gaussian mixture models \cite{plataniotis2017gaussian}, due to their  universality at approximating densities \cite{Goodfellow-et-al-2016} and because of their extensive use in many practical settings. Yet, it becomes evident that, as far as its ``off-the-shelf'' performance, it requires fine-tuning specific to the problem at hand. While the statement might appear obvious, it scratches the surface of what is the additional degree of fine-tuning necessary to produce good results that balance modeling effort (few training data and parsimony in terms of parameterization) for performance (seen as accurate reconstruction of distributions in the MSE or GKL-divergence sense). We consider alternatives, where the Gaussian mixture is abandoned in favor of a distribution that better fits the surface shape of the target distribution, namely the M\"obius distribution.

As an alternative to standard mixture models, we also investigate the use of NFs \cite{papamakariosNormalizingFlowsProbabilistic2021} to approximate the spatial density. NFs learn an overall invertible, differentiable transformation---typically constructed by composing a sequence of simpler bijections---that maps a simple, tractable base distribution into a more complex target distribution. Here, the standard approach is to start with a Gaussian distribution but sufficiently ``distort'' it via transformations. The general expectation is that, NFs require more computational effort in training but might win in terms of accuracy. 

Ideally, for a machine learning model to be convincing, it is necessary that it is also interpretable by human experts. The example of Figure \ref{fig:exact} illustrates a criterion we could use for such interpretability. For example, a model that captures the background centered dome in one component distribution, and the off-center peak in another component can be claimed to be  more interpretable than one that does not provide such separation.  Hence, NF models may indeed be performing well but suffer from poor (or complete lack) of interpretability. 

The contributions of the current paper are all related to learning spatial distributions by means of MDNs and NFs and have the characteristics of: 
\begin{itemize}
    \item capturing distributions defined on the disk, as required by many existing mobility models, 
    \item capturing distributions that include ``features'' such as local attraction points/peaks,
    \item requiring a small number of training set data to produce a good model in terms of parsimony and prediction accuracy, and 
    \item resulting in interpretable results.
\end{itemize}

In Section \ref{sec:pd} we provide a specific problem definition from the point of view of approximating a particular variety of density functions derived from nodemobility. The point is that a wider set of similar distributions can be handled using similar models.  Section \ref{sec:models} introduces the family of M\"obius distributions in a mixture setting analogous to mixture-of-Gaussians. It provides motivating examples of how the M\"obius approach intrinsically enforces boundary conditions and captures complex density shapes, ultimately outperforming both Gaussian mixtures and standard normalizing flows. The summary of learning results is given in Section \ref{sec:eval}. %
Section \ref{sec:related} links our work with a number of current trends in learning models comparing and contrasting them. Finally, Section \ref{sec:conclusions} summarizes the paper and points to a generalization toward stochastic model parameterization.

\section{Background \& Problem Definition}\label{sec:pd}

When engineering systems, such as communication systems, it is convenient to have closed-form formulations for relevant probability density functions and/or their related statistical properties. For example, the closed-form relation between the average delay and arrival rates at a queueing system, as per Little's law, is often employed in solving bigger network design problems \cite{9155442}. \textit{Here we are interested in serving the needs of spatial optimization problems, i.e., placement of resources, where the placement impacts a probability density function.} Such optimization problems arise e.g., in the study of node density of mobile nodes moving around under certain mobility rules. Part of the assumptions when studying such spatial densities is how far is the extent of the space over which the distribution should be described. Motivated by the fact that many applications involve the placement of resources in cities and other population centers, an often applied assumption is that the area is circumscribed by a circle. Hence, densities over a disk are commonplace \cite{yan2017universal}.

A well-studied baseline mobility model is the RWP model, which results in a closed-form probability density function of nodes across space \cite{hyytia2006spatial}. RWP assumes that nodes continuously move around along straight paths from waypoint to next waypoint. Yet, when RWP is extended to capture more subtle movement rules, e.g., movement predicated on the state of the mobile nodes, and non-uniformity of space, extracting a closed-form solution is challenging and largely unsuccesful. For example, RWP alone does not have a concept of points of attraction. Attraction points could be inflicting a change in the mobility behavior depending on the state and proximity of the mobile nodes. A case of this behavior arises when placing charging stations at specific locations and, as a result, attracting users to go to them to charge their mobile devices. One expects the density of mobile nodes to be higher around chargers. The less the overall available energy, the more the likelihood they will be attracted to the charger and the density will ``bulge'' around the charger which will influence their trip between successive waypoints. Deriving a function to describe the density based on pertinent parameters, e.g., location of the charger, average energy capacity of nodes, etc., is necessary in such an application. 

In previous work \cite{9678073}, we have derived the node density distribution for a network of energy-constrained mobile nodes. We are repeating here some of the key formulas behind the analytical expression of the node density to illustrate their intertwined nature. Note that we  have thus far found \textit{no closed-form solution} for the distribution. Additionally, it is computationally expensive to produce a numerical approximation to the solution, in particular due to the  multiple integrals involved. The purpose of the learning described in the next section is to learn an efficient function that evaluates to this complicated density using a \textit{small training set} since each training point requires substantial numerical processing or lengthy simulations.

The density is a complicated one, owing to the fact that it models chargers that ``attract'' nodes whose energy is depleted. The trajectory of the mobile nodes is altered, from continuing to their destinations, to diverting to a charger before proceeding to their destinations. The decision to divert is only taken if the mobile node is a certain, relatively small, distance from the charger---equivalently seen as the ``attraction'' range of a charger. More formally, suppose that $c=(r_c, \theta_c)$ (our first two parameters) denotes the charger's location in polar coordinates. That is, $r_c$ and $\theta_c$ are radial and angular coordinates respectively, where the radius of the disk (centered at (0,0)) is normalized to a unit (hence $r_c<1$).  The (energy-depleted) mobile node will divert for recharging, provided that its current location, denoted by $z=(x,y)=r(\cos\vartheta, \sin\vartheta)$, is within the (closed) disk, $\overline{\mathcal{D}}(c, d')=\left\{z\mid |z-c|\leq d'\right\}$, i.e., if it is within $d'>0$ distance from the charger ($d'$ being the third parameter of our model). The limitation of attracting a depleted node only if it is within an attraction range models the reluctance of a node to divert significantly from their path to the destination for the sake of recharging (which they can do at their destination point anyway). 

Besides adequate proximity to the charger, the other premise of the diversion for recharging, i.e., that the node is depleted, is made tantamount to the event that the node has traveled a distance $d>0$ ($d$ is our fourth parameter) after departure  from its last waypoint. This projection of energy consumption processes onto distance is a modeling convenience allowing us to lump together the effect of many factors (e.g., rated power, moving speed) into a typical ``energy budget'' expressed as a distance threshold $d$, in the same units as those for determining the attraction ($d'$) to the charger. In this example application, an implicit assumption is that the energy for locomotion is separate from the energy for computation and communication---a scenario which fits the case of human movement with portable devices (e.g., smartphones, wearables) being carried around.

Even in the presence of a single charger, the formula we derived for the probability density is non-trivial and has so far evaded a closed form expression \cite{9678073}, leading to needs for numerical evaluation. Note that we have been able to derive closed form expressions for the 1-dimensional case \cite{8653318}, but its components and constraints are significantly less challenging than the 2-dimensional case. The following formula of node density for the 2-dimensional case applies when $r_c<1-(d+d')$, i.e., when the charger is placed relatively close to the center to allow nodes that start moving from the boundary to have a chance to be attracted to the charger when their energy is depleted.  The high-level structure of the spatial density $f_{Z(t)}(z)$ satisfies the following equation:
\begin{equation}\label{eq:f_z}
 \pi^2\alpha\cdot f_{Z(t)}(z)
 =f_{\mathrm{RWP}}(z) + \begin{cases}
  g_0(z) - h_0(z),&\mathrm{if}\ z\in\Omega\setminus\overline{\mathcal{D}}(c, d')\\
  g_1(z) - h_1(z),&\mathrm{if}\ z\in\overline{\mathcal{D}}(c, d')\setminus\{c\} %
 \end{cases}
\end{equation}
where $\alpha$ denotes the expected distance (including diversions to the charger) between any two waypoints, $f_{Z(t)}(z)$ is the (stationary) probability density of the node's location $Z(t)$ over time $t$, and $f_{\mathrm{RWP}}(z)$ is the density assuming only RWP mobility (i.e., with no diversion to the charger).

Note that the movement area and thus the support of $f_{Z(t)}(z)$ is a \emph{unit disk} $\Omega=\{z\in\mathbb{R}^2\mid\left|z\right|\leq 1\}$. The density within the attraction range of the charger (i.e., for $z\in\overline{\mathcal{D}}(c, d')$) can be captured, excluding the basic RWP component $f_{\mathrm{RWP}}(z)$ (and the coefficient $(\pi^2\alpha)^{-1}$), by an infinity at the charger (i.e., for $z\in\{c\}$) and a subtraction of two non-trivial functions $g_1(z)$ and $h_1(z)$; a similar superimposition onto $f_{\mathrm{RWP}}(z)$, with the subtraction $g_0(z)-h_0(z)$, is also found for formulating the nodal density beyond the attraction range (for $z\in\Omega\setminus\overline{\mathcal{D}}(c, d')$). The respective forms of $g_1(z)$, $h_1(z)$, $g_0(z)$, and $h_0(z)$ are given by
\begin{empheq}[left={\empheqlbrace}]{alignat*=2}
 g_1(z)%
 =&\int_0^{d'-|z-c|} \left(1+\frac{l_{16}}{|z-c|}\right)\cdot\frac{\gamma_1}{2} \dd{l_{16}} \\
  & + \int_{d'-|z-c|}^{l_{17}} \left(1+\frac{l_{16}}{|z-c|}\right)\cdot\frac{\gamma_0}{2} \dd{l_{16}}\numberthis\label{eq:g_1}\\
  & + \int_{\theta_5+\frac{\pi}{2}}^{\theta_5+\frac{3\pi}{2}} \frac{d'}{|z-c|}\cdot\frac{\varphi_2}{2} \dd{\theta'_0},\\
 h_1(z)=&\ 4(1-r^2)E(r)-2d^2 E(r)-\pi d,\numberthis\label{eq:h_1}\\
 g_0(z)%
 =&\int_0^{l_{17}} \left(1+\frac{l_{16}}{|z-c|}\right)\cdot\frac{\gamma_0}{2} \dd{l_{16}},\numberthis\label{eq:g_0}\\
 h_0(z)=&\int_{\theta_5-\theta_6}^{\theta_5+\theta_6} \frac{\varphi_0}{2} \dd{\theta'_0}.\numberthis\label{eq:h_0}
\end{empheq}
where
\begin{empheq}[left={\empheqlbrace}]{alignat*=4}
 \gamma_1=&\int_0^{2\pi} l_{12}(l_{12}+2d) + (l_{13}-d)(l_{13}+d) \dd{\theta'_0},\numberthis\label{eq:gamma_1}\\
 \gamma_0=&\int_{\theta_5-\theta_8}^{\theta_5+\theta_8} (l_{13}-l_{14}-d)(l_{13}+l_{14}+d) \dd{\theta'_0},\numberthis\label{eq:gamma_0}\\
 \varphi_2=&\ -\cos(\theta'_0-\theta_5)\cdot \hat{l}_{12}(\hat{l}_{13}-d)(\hat{l}_{12}+\hat{l}_{13}+d),\numberthis\label{eq:varphi_2}\\
 \varphi_0=&\ l_2 (l_3-l_4-d)(l_2+l_3+l_4+d),\numberthis\label{eq:varphi_0}
\end{empheq}
and where $l_2$, $l_3$, $l_4$, $l_{12}$ ($\hat{l}_{12}$), $l_{13}$ ($\hat{l}_{13}$), $l_{14}$, $l_{16}$, and $l_{17}$ are all lengths of line intervals between the node's location (or, its current destination) and a certain intersection point along the node's moving direction, with or without recharging at the charger. For instance, $l_2$ is the distance from the node's location $z$ straight to the boundary of the (disk-shaped) movement area along its moving direction (at angle $\theta'_0$), in the case of no need (or no opportunity) for recharging, while $l_{16}$ is the distance from $z$ to its original destination $w_1$ assuming that the node just departed from the charger, after recharging, for $w_1$. Detailed illustration of the relevant line segments and other variables can be found in \cite{9678073}. Nevertheless, the density function given by Equation \ref{eq:f_z} is too complicated to derive a closed-form expression for it. In fact, it might be computationally more convenient to perform simulations than attempt a numerical solution altogether! Yet, both simulation and numerical evaluation are undesirable when the density function is just a subroutine repetitively called  from a larger optimization engine to solve a spatial placement problem.

The question we are trying to answer is, \emph{what learning technique could we employ to derive, from a small training set, an accurate node density distribution across the domain of interest (a disk in our case)?} The parameters of this function will be the  $d$, $d'$, $r_c$ and $\theta_c$ as presented above.  If indeed such a form is learned, e.g.,  as a mixture of component distributions, we could also, depending on the learning technique used, tease out its constituent components, to gain further insights to the density distribution  \cite{9678073}, especially if the components come from elementary well-understood functions, e.g., from Gaussians. To this end, we next focus on putting the research question in the context of machine learning.

\section{Mixture Networks and Normalizing Flows} \label{sec:models}

\begin{figure*}[!t]
      \centering
      \includegraphics[width=6.9in]{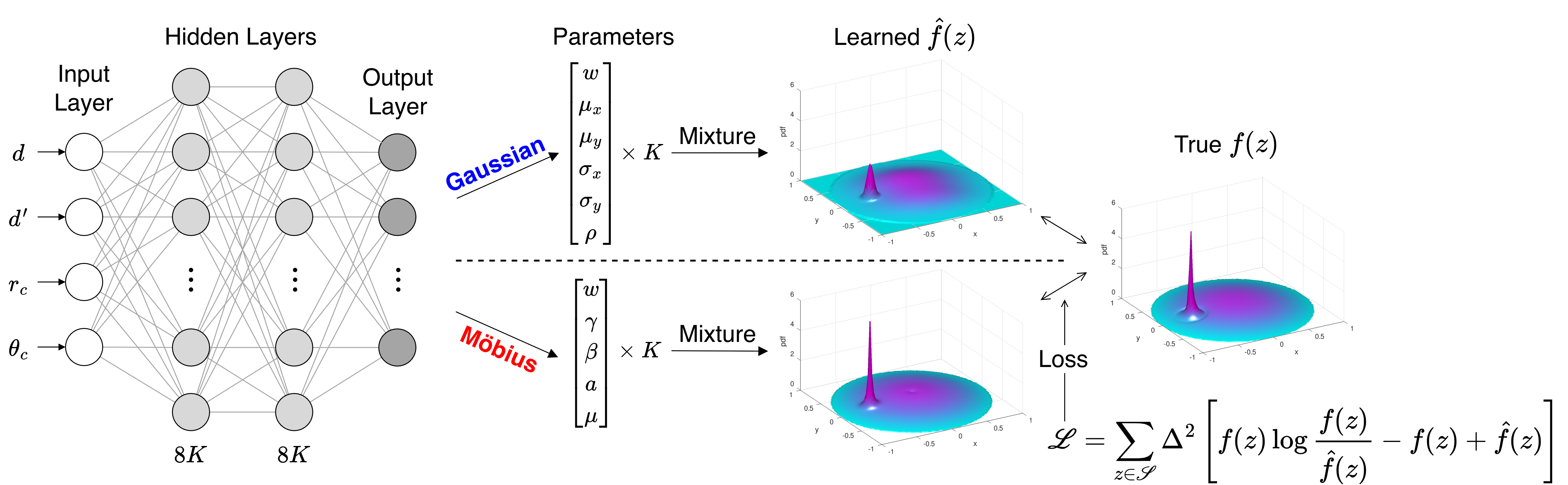}
      \caption{Overview of the deep learning models using mixtures of M\"obius versus Gaussian distributions.}
      \label{fig:fnn}
     \end{figure*}

The general framework of density learning based on mixtures of Gaussian (or M\"obius for that matter) distributions can be found in Figure \ref{fig:fnn}, which is inspired by the MDN first introduced in \cite{bishop1994mixture}. The basic idea of MDNs is to approximate an unknown probability distribution conditioned on the input, through a weighted sum of $K\in\mathbb{Z}_{\geq 1}$ distributions that are more manageable.

Note that Gaussian distributions are the most commonly assumed kernels in MDNs, since it has been proved that any density function in question can be approximated as accurately as possible by a Gaussian mixture model that is properly parameterized \cite{mclachlan1988mixture}. Despite the high flexibility and generality of Gaussian distributions, however, it remains an open question as to whether and how new distributions can be used to be better tailored (e.g., at a lower cost) to the needs of case-specific density approximations. In this paper, we answer the question of matching the finitely-supported density observed in (charging aware) mobile nodes by focusing and further elaborating on the (beta type \Rn{3}) M\"obius distribution, which is introduced to substitute for the Gaussians. %

As seen in Figure \ref{fig:fnn}, the input to the learning models is the 4-tuple characterizing an instance, i.e., $\lambda=(d, d', r_c, \theta_c)$, corresponding to the energy availability ($d$) of the node and the radius ($d'$) and location ($r_c$, $\theta_c$) of the charger (we start with the single-charger case). Through a fully connected deep neural network (containing two hidden layers of size $8K$ in the example), a number of $6K$ parameters, for Gaussian distributions (see Section \ref{subsec:gauss}), or $5K$ parameters, for M\"obius distributions (see Section \ref{subsec:mob}), are produced at the output layer.

Regardless of whether we train for Gaussian or M\"obius mixtures, the training loss function (Equation (\ref{eq:loss})) is evaluated over a set of density values sampled from the continuous derived distribution, at coordinate points defined by the set $\mathcal{S}$. While we are free to define set $\mathcal{S}$ arbitrarily, our work is sampling uniformly on the x- and y- axis, %
with a step size $\Delta = 0.02$.
Only points within the circular support are of relevance to the density while points outside of it have zero density; correspondingly, the coordinate points in $\mathcal{S}$ are also limited to those within the circular support, i.e., each point in $\mathcal{S}$ having a radius no larger than 1.\footnote{Specifically, $\left|\mathcal{S}\right|=7860$.} For the loss function, we employ the \emph{generalized} Kullback-Leibler (KL) divergence \cite{kullback1959information}, evaluated on the discrete probability masses obtained by multiplying each density value by the cell area $\Delta^2$ (here, $\Delta^2=0.02^2$), i.e.,
\begin{equation}\label{eq:loss}
\mathscr{L} = \sum_{z\in\mathscr{S}} \left[ P(z)\log\frac{P(z)}{Q(z)} - P(z) + Q(z) \right],
\end{equation}
with $P(z)=f(z)\,\Delta^2$ and $Q(z)=\hat{f}(z)\,\Delta^2$. Here, $f(z)$ is the ground truth acquired either by simulation or by numerical evaluation of the density in Equation (\ref{eq:f_z}), while $\hat{f}(z)$ denotes the density obtained by the learning models. The generalized KL (GKL) divergence is adopted, because it defines a proper, non‑negative training loss over $\mathcal{S}$, even when the total masses of $P(z)$ and $Q(z)$ differ over $\mathcal{S}$ (e.g., $\sum_{z\in\mathcal{S}} Q(z) < \sum_{z\in\mathcal{S}} P(z) = 1$). When those totals agree, it reduces to the traditional KL divergence. %

\subsection{Mixture of Gaussians and Its Limitations}\label{subsec:gauss}
As illustrated in Figure \ref{fig:fnn}, when (bivariate) Gaussian distributions are used as the kernels, the output layer of the neural network is of size $6K$, including $K$ mixing weights ($w$), $2K$ means ($\mu_x$, $\mu_y$), $2K$ standard deviations ($\sigma_x$, $\sigma_y$), and $K$ correlation coefficients ($\rho$). Although Gaussians are popular for mixture for mobility modeling \cite{alessandretti2020scales, tang2019estimating}, the application of such classic distributions to our case has deficiencies that are critical to the efficacy of density approximation, as elaborated below.

1) \emph{Infinite support vs. encircled movement area}. Gaussian distributions have infinite support, and this trait is inherited by their mixtures,  causing non-zero probability to ``leak'' out of the bounded area. Numerically, our experiments (see Section \ref{sec:eval} for details and more results) show that the fully trained Gaussian mixture model can leak a probability mass from 1.45\% to 7.63\% (depending on the value of $K$) outside the disk area for the training set. When the mixture model is applied as a subroutine, such leakage may considerably disrupt downstream optimization algorithms; for instance, attempting to naively mask the leakage via rejection sampling can introduce unpredictable computational overhead and break the differentiability required by gradient-based solvers.

One may consider truncating the distributions for a finite support \cite{azam2019bounded}, yet it is often non-trivial to analytically evaluate the normalizing factors for truncation. There exists no closed form for finding the exact truncated volume of an arbitrarily generated bivariate Gaussian over the unit disk (which is the finite support we assume). Alternatively, one may resort to numerical integration for approximate normalization, which, however, would require repeated two-dimensional numerical integration during training. Besides, as shown in Figure \ref{fig:exact}, the density distribution of the mobile node is characterized by continuously zero-diminishing values approaching any point on the border of the disk, and it takes effort to keep this feature while imposing truncation on the Gaussians, be it exact or numerical for normalization. 
Artificially severing Gaussians at the boundary creates unnatural hard edges and discontinuities, which again can impact the smooth gradients relied upon by downstream algorithms.

\begin{figure*}[!t]
 \centering
 \subfigure[]{
 \label{fig:gau_bvm_mob:a}
 \includegraphics[width=5.6cm]{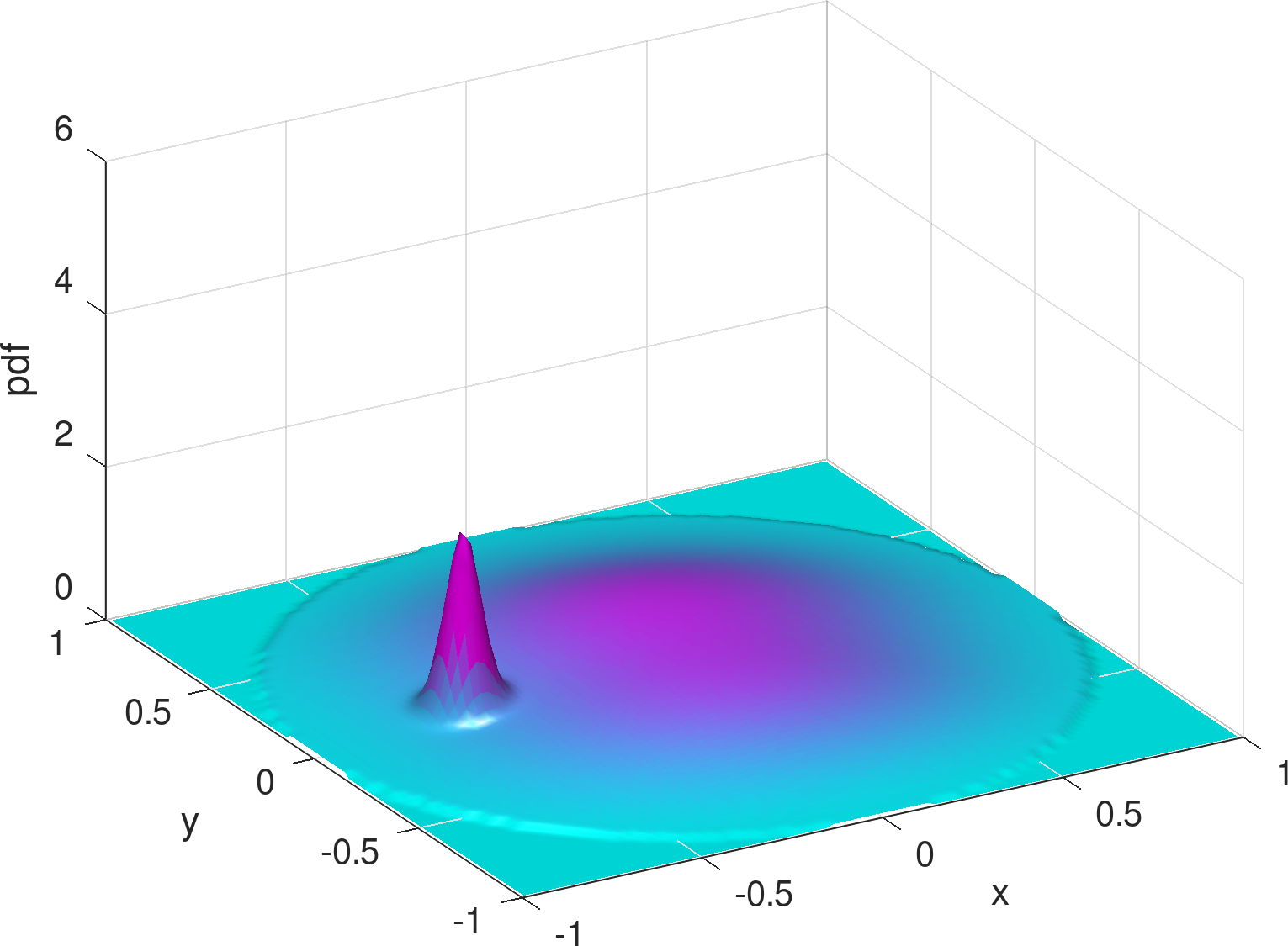}}
 \subfigure[]{
 \label{fig:gau_bvm_mob:b}
 \includegraphics[width=5.6cm]{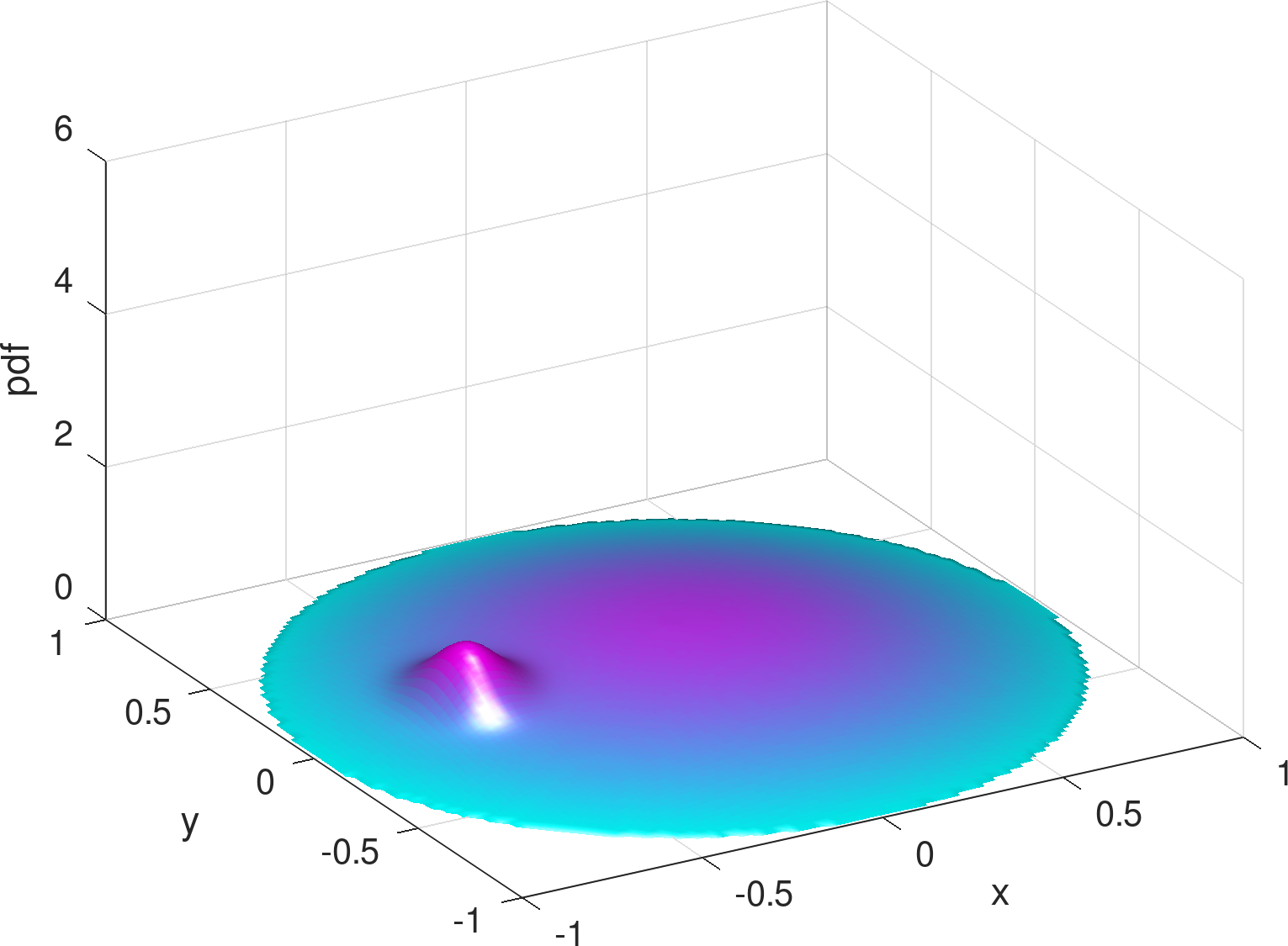}}
 \subfigure[]{
 \label{fig:gau_bvm_mob:c}
 \includegraphics[width=5.6cm]{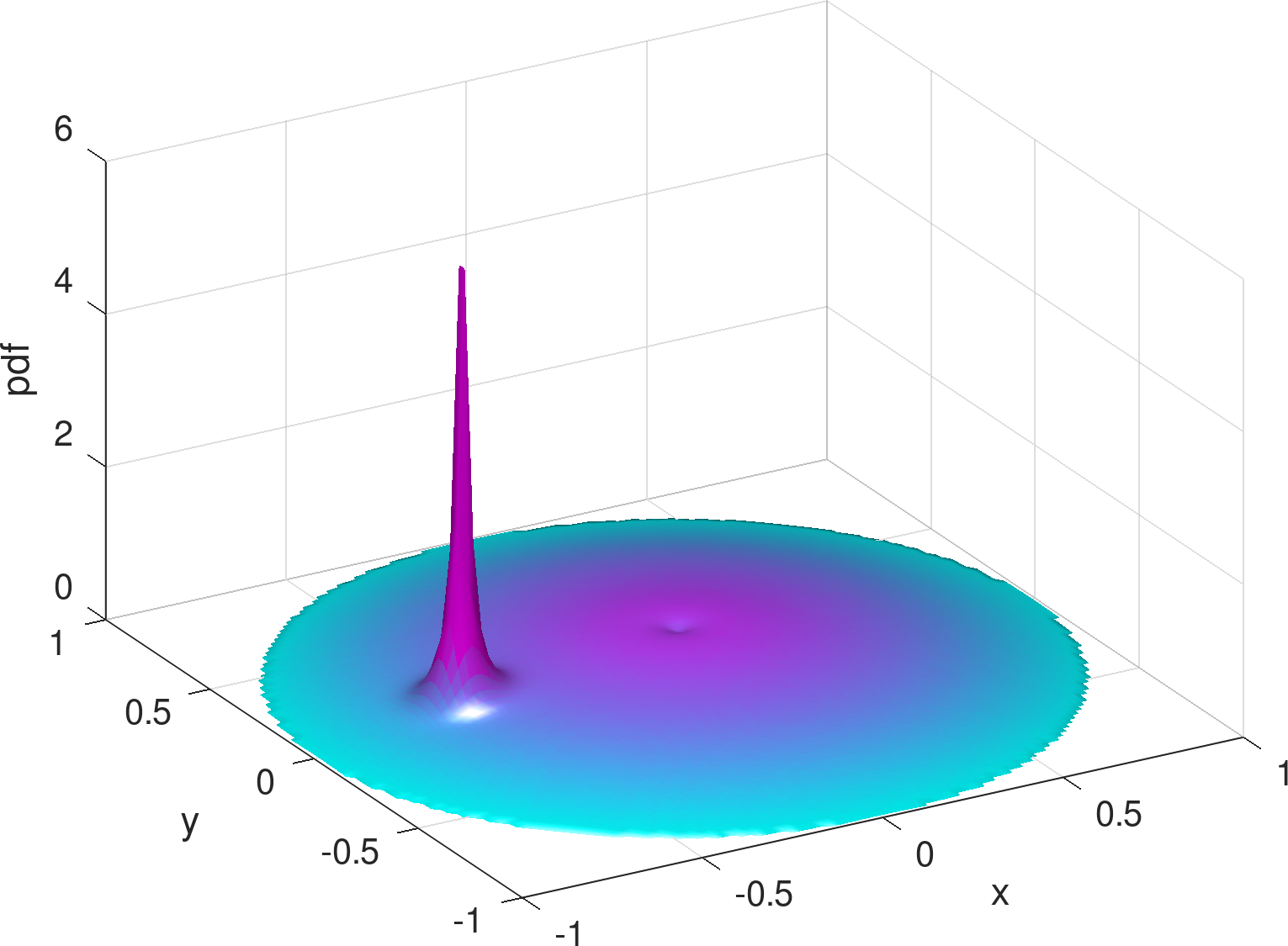}}
 \subfigure[]{
 \label{fig:gau_bvm_mob:d}
 \includegraphics[width=5.6cm]{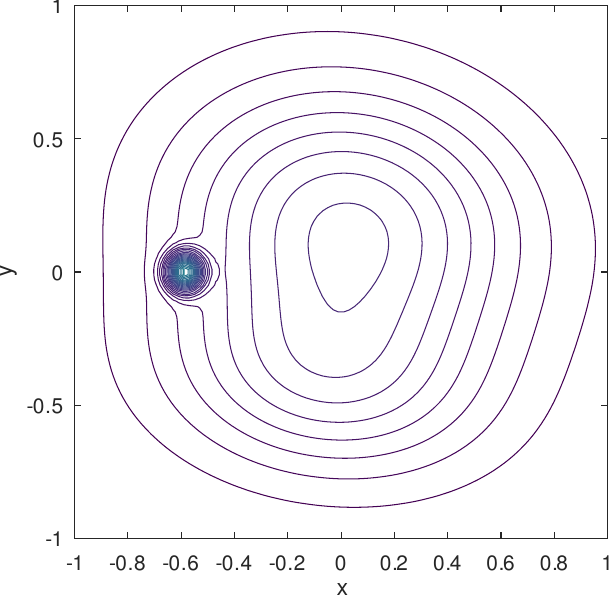}}
 \subfigure[]{
 \label{fig:gau_bvm_mob:e}
 \includegraphics[width=5.6cm]{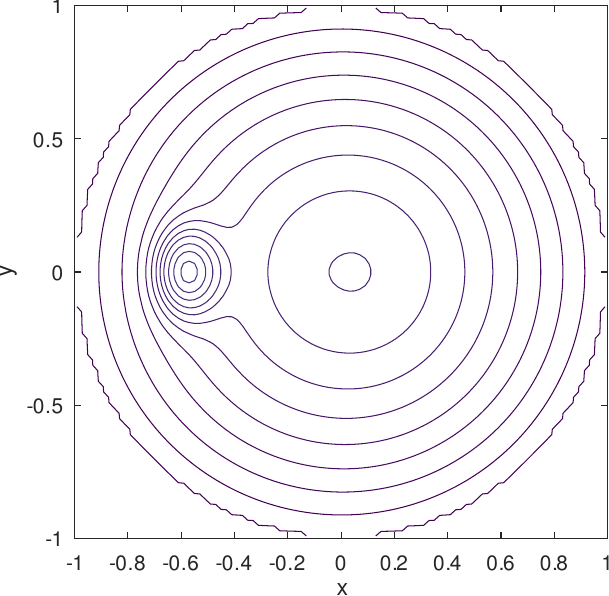}}
 \subfigure[]{
 \label{fig:gau_bvm_mob:f}
 \includegraphics[width=5.6cm]{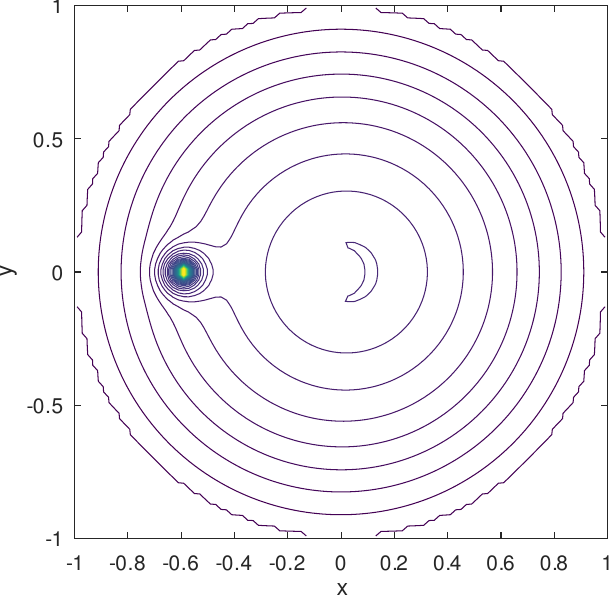}}
 \caption{Mixtures of three component distributions (i.e., $K$=3), plotted from a training set of 24 points: (a)(d) Gaussian as the kernel, with \(\text{MSE}=0.00549\) and \(\text{GKL}=0.01901\); (b)(e) Beta-von-Mises as the kernel, with \(\text{MSE}=0.00969\) and \(\text{GKL}=0.01010\); (c)(f) M\"obius as the kernel, with \(\text{MSE}=0.00064\) and \(\text{GKL}=0.00109\).}
 \label{fig:gau_bvm_mob}
\end{figure*}

2) \emph{Bell surface vs. concave} $f_{\mathrm{RWP}}(z)$. The shapes of bivariate Gaussian distributions are subject to a particular bell surface, which is unnatural and inconvenient for fitting distributions that are radically different in shape. For illustration, Figures \ref{fig:gau_bvm_mob:a} and \ref{fig:gau_bvm_mob:d} exemplify the output of the learning model from Figure \ref{fig:fnn} (see Section \ref{sec:eval} for the details), assuming three Gaussians are mixed (i.e., $K=3$). In comparison to the exact density distribution in Figure \ref{fig:exact}, the fully trained Gaussian mixture model captures the accumulation at the charger location only at a lower height and appears structurally ill-suited to mimicking the RWP component, $f_{\mathrm{RWP}}(z)$, which is spherically concave.

To delve into the reason, we can inspect the three Gaussian components separately, as shown in Figures \ref{fig:g_m} (a)--(f). It can be seen that minimization of the loss function of Equation (\ref{eq:loss}) gives rise to a scattering of two Gaussians over the movement area for fitting $f_{\mathrm{RWP}}(z)$, besides one Gaussian used to approximate the peak around the charger. This is not surprising, since the Gaussian distribution is relatively poor fit for distributions like in Figure \ref{fig:exact}. On the one hand, a single Gaussian lacks the flexibility to fully approximate the extremely concave cone shape around the charger. On the other hand, each Gaussian is bell-shaped and thus needs multiple other Gaussians to be placed apart, on its outskirts, to complement the convexity around its tail areas for a finer approximation. When the number of components is three, with only two (the other one needed for the peak) to fit $f_{\mathrm{RWP}}(z)$, the convexity around either Gaussian component cannot be fully covered, resulting in the asymmetric, stretched dome in Figures \ref{fig:gau_bvm_mob}(a)(d).

3) \emph{More parameters vs. low-cost learning}. A direct remedy for the limitation of bell-shaped Gaussians for approximating $f_{\mathrm{RWP}}(z)$ is to simply increase the number of components to be mixed. However, as we will see in Section \ref{sec:eval} (e.g., Table \ref{tab:g_vs_m}), %
the gap in shape (in terms of MSE and GKL) between the learned Gaussian mixture and the exact density distribution remains relatively wide even for $K$ increased up to 10 (i.e., 60 parameters to learn in total), while it entails scaling up the neural network proportionally to accommodate more parameters for output (noting that each hidden layer is of size $8K$).

Another potential solution is to modify the Gaussian distribution somehow for better fitness. If one considers generalizing Gaussians for greater malleability, however, it can be costly (in terms of the number of parameters) since more additional parameters need to be introduced into each Gaussian component \cite{arnroth2023some}, adding to the parameters to estimate in total. On the other hand, it is non-trivial (if doable) to devise a specialized but simpler variant of the Gaussian distribution for our density fitting problem, while the susceptibility to an unbounded support still needs to be addressed. Hence, rather than tackle the conundrum faced by Gaussian distributions, it may be easier to find another suitable type of distribution that costs less for mixture.

\subsection{Mixture of Beta-von-Mises Distributions}
A logical intermediate step to address the limitations of the Gaussian mixture model is to adopt a distribution that is inherently finite-supported. Using polar coordinates \( (r, \vartheta) \), one can assemble a radial-angular separable mixture kernel, where the radial distance \( r \in [0,1] \) is modeled by a beta distribution while the angle \( \vartheta \in (-\pi, \pi] \) by a classic von Mises distribution \cite{mardia2000directional}. Expressed in polar coordinates (requiring a Jacobian of $\frac{1}{r}$ if transformed to Cartesian), the density of a single beta-von-Mises component over the disk can be given by
\begin{equation}\label{eq:bvm}
f(r,\vartheta) = f_{\text{beta}}(r; \alpha_r, \beta_r) f_{\text{vM}}(\vartheta; \mu_{\vartheta}, \kappa),
\end{equation}
where \( \alpha_r, \beta_r > 0 \) jointly control the radial shape, \( \mu_{\vartheta} \in (-\pi, \pi] \) is the mean angle, and \( \kappa > 0 \) controls the angular concentration. 

This beta-von-Mises kernel directly addresses the aforementioned drawbacks of the Gaussians. It eliminates the leakage problem, strictly confining the probability mass to the unit disk without requiring artificial truncation. Also, it is highly parameter-efficient, requiring only four parameters (i.e., \( \alpha_r, \beta_r, \mu_{\vartheta}, \kappa \)) per component, which is fewer than the five required for a bivariate Gaussian. %

However, although it allows evading the rigid bell surface of a Gaussian, the beta-von-Mises kernel imposes the built-in constraint of strict independence between the radial and angular factors (Equation (\ref{eq:bvm})). Consequently, while it can easily model shapes like origin-centric sectors, it struggles to represent circular, narrow peaks off the center. As illustrated in Figures \ref{fig:gau_bvm_mob:b} and \ref{fig:gau_bvm_mob:e}, while the beta-von-Mises mixture can reproduce the symmetric RWP dome $f_{\mathrm{RWP}}(z)$ more faithfully than Gaussians, the charger peak fitted by the former spreads into a broad, flattened footprint, with a height even lower than that of the peak yielded by Gaussians. Fixing this smearing would require mixing a larger number of components, cancelling the low-cost advantage of the beta-von-Mises kernel. This motivates the need for a distribution that is bounded, parsimonious, and capable of natively modeling joint spatial correlations.

\subsection{Mixture of M\"obius Distributions}\label{subsec:mob}

To overcome the aforementioned deficiencies of both the Gaussian and beta-von-Mises mixture models, we make use of, as the kernel for mixture, the so-called beta type \Rn{3} M\"obius distribution \cite{sym11081030}, which is an extended variant of the (basic) M\"obius distribution originally proposed in \cite{jones2004mobius}. Like the beta-von-Mises kernel, the first key merit of this distribution is that it has a support of $0\leq x^2+y^2 \leq 1$ (in Cartesian coordinates), i.e., a unit disk, which exactly fits the support of the node density introduced in Section \ref{sec:pd}. Specifically, the beta type \Rn{3} M\"obius distribution can be formulated as follows:

\begin{equation}\label{eq:mob_c}
\begin{split}
f(x,y) =&\; C (1-a^2)^{\gamma+1}(1-x^2-y^2)^{\gamma-1}\\
& \times \left[a^2-2ax\cos\mu-2ay\sin\mu+(x^2+y^2)\right]^{\beta-1}\\
& \times \left[(1+x^2+y^2)(1+a^2)-4ax\cos\mu-4ay\sin\mu\right]^{-\gamma-\beta},
\end{split}
\end{equation}
where
\begin{equation}
C=\frac{2^{\beta}\Gamma(\beta+\gamma)}{\pi\Gamma(\beta)\Gamma(\gamma)},
\end{equation}
is the normalizing constant (with $\Gamma(\cdot)$ being the gamma function), and the four parameters, $\gamma\in(0,+\infty)$, $\beta\in(0,+\infty)$, $a\in[0, 1)$, and $\mu\in(-\pi, \pi]$, control the steepness of concentration, modality (i.e., bimodality when $\beta>1$ and unimodality otherwise), skewness/asymmetry, and orientation of the distribution, respectively.

It is clear that the tuning of a beta type \Rn{3} M\"obius distribution only involves four parameters, one parameter less than a Gaussian distribution. This is the second key merit of the M\"obius distribution, i.e., having a relatively low cost of learning (in terms of the number of parameters), for a given value of $K$, matching the efficiency of the beta-von-Mises kernel. The third key merit of the M\"obius-based mixture model, which may be less obvious, is its ability to emulate the shape of the node density that we are concerned with, despite fewer parameters used. %

As a hint, Figures \ref{fig:gau_bvm_mob:c} and \ref{fig:gau_bvm_mob:f} show the distribution mixed from three beta type \Rn{3} M\"obius distributions that are sufficiently tuned by the learning model. Firstly, it is able to capture the unique charger peak accurately with only one component (see Figures \ref{fig:g_m}(g)(j)), sparing two other components to jointly depict the RWP dome (Figures \ref{fig:g_m}(h)(i)(k)(l)). Secondly, the RWP component learned by the remaining M\"obiuses retains the concavity (although with a minor dimple at the origin) and symmetry of the exact shape in Figure \ref{fig:exact}. Hence, the M\"obiuses can overcome the shortcomings of both the mixtures of Gaussians, which are unbounded and unsuited for fitting the RWP dome, and of beta-von-Miseses, which tend to underfit the peak at the charger. The superiority is confirmed by the numerical results, i.e., the GKL of the exact density distribution (Figure \ref{fig:exact}) from the M\"obiuses can decrease to a low value of 0.00109, far lower than the value of 0.01901 from the Gaussians and 0.01010 from the beta-von-Miseses. The M\"obiuses are also better in terms of MSE, with $0.00064$, which is lower than both $0.00549$ (Gaussians) and $0.00969$ (beta-von-Miseses).

\subsection{Applying Normalizing Flows}

As an alternative to MDN, we consider Normalizing Flows (NFs). We follow the standard approach to implementing NFs \cite{papamakariosNormalizingFlowsProbabilistic2021}. Given a spatial coordinate vector \( z=(x,y) \) and a condition vector \( \lambda = (d, d', r_c, \theta_c) \), the density is evaluated using the change of variables formula:
\begin{equation}
f(z \mid \lambda) = p_U(F(z; \lambda)) \left| \det \frac{\partial F(z; \lambda)}{\partial z} \right|,
\end{equation}
where \( F(z; \lambda) \) is this overall learned bijective transformation, \( p_U \) is the base distribution of the latent space, and \( \smash{ \left| \det \frac{\partial F(z; \lambda)}{\partial z} \right| } \) is the absolute determinant of the Jacobian of \( F(z; \lambda) \).

To evaluate the efficacy of NFs in capturing the bounded, non-trivial node density, we implement two baseline models (i.e., RNVP and NSF, as to be elaborated later) that share a common coupling construction and differ only in their transformation and base distribution. To better respect the circular geometry of the problem, both models operate on the polar coordinates \( (r, \theta) \) of the grid points, and \( F(z; \lambda) \) is composed of a sequence of \( L \in \mathbb{Z}_{\geq 1} \) coupling layers. In the \( n \)-th layer (\( n = 1, \dots, L \)), the input vector \( h^{(n-1)} \) (with \( h^{(0)} \) being the initial polar coordinates) is partitioned into two halves, \( h^{(n-1)}_1 \) and \( h^{(n-1)}_2 \), and the forward transformation to the output \( h^{(n)} = (h^{(n)}_1, h^{(n)}_2) \) is defined as:
\begin{equation}\label{eq:coupling}
\begin{cases}
h^{(n)}_1 = h^{(n-1)}_1 \\
h^{(n)}_2 = g^{(n)}\!\left( h^{(n-1)}_2 ;\, \phi^{(n)}(h^{(n-1)}_1, \lambda) \right),
\end{cases}
\end{equation}
where \( g^{(n)}(\cdot; \cdot) \) is a monotonic, invertible transformation of the second half, with parameters emitted by a conditioner network dedicated to that layer, i.e., \( \phi^{(n)}(\cdot, \cdot) \), which takes the (untransformed) first half and the condition \( \lambda \) as its inputs. Each conditioner comprises of three hidden layers of 64 neurons each.\footnote{This capacity is determined empirically as an optimal balance between sufficient expressiveness and excessive complexity. We observed that doubling it to 128 neurons yielded only marginal performance gains while evidently introducing extra computational costs.} 
Because \( h^{(n)}_1 \) is passed through unchanged, the Jacobian of each layer is triangular and its determinant reduces to the derivative of \( g^{(n)}(\cdot; \cdot) \). Successive layers alternate the roles of the two halves \cite{dinh2017density}, so that each coordinate is eventually transformed as layers are stacked.

Our first baseline NF model is the well-known real non-volume preserving (RNVP) architecture \cite{dinh2017density}, for which \( g^{(n)}(\cdot; \cdot) \) is affine, i.e.,
\(h^{(n)}_2 = h^{(n-1)}_2 \odot \exp(s^{(n)}) + t^{(n)}\), with \( (s^{(n)}, t^{(n)}) \) being the scale and translation outputs of \( \phi^{(n)}(\cdot, \cdot) \). The Jacobian determinant of the \( n \)-th layer is then simply \( \exp(s^{(n)}(h^{(n-1)}_1, \lambda)) \). The final transformed coordinates \( h^{(L)} \) are mapped back to the Cartesian space to evaluate the probability density of the base distribution \( p_U \), incorporating a Jacobian (i.e., the ratio of the final to initial radial coordinates) accounting for the coordinate transformation into the overall determinant. We parameterize \( p_U \) as a single-component Gaussian distribution, where the means and variances are dynamically learned by a feed-forward neural network conditioned on \( \lambda \), consisting of two hidden layers of 8 neurons, similar to the Gaussian mixture model in Figure \ref{fig:fnn} with $K=1$. While RNVP is highly flexible, this Gaussian base still possesses an infinite support, hence being prone to the same leakage issue and requiring truncation to confine the density.

To avoid truncation, our second NF baseline keeps the density confined such that the support matches the disk area natively. Following the neural spline flow (NSF) approach \cite{durkan2019neural}, \( g^{(n)}(\cdot; \cdot) \) in Equation (\ref{eq:coupling}) is constructed as a rational-quadratic spline acting on the interval natural to the polar coordinates being transformed, i.e., \( r \in [0,1] \) and \( \theta \in (-\pi, \pi] \). The conditioner accordingly emits the widths, heights, and inner-knot derivatives of $K_b$ spline bins ($K_b=8$ here) rather than a scale and a translation, and each layer is initialized to the identity map for training stability. Since every spline maps its interval onto itself, the transformed coordinates remain valid polar coordinates of the unit disk, which allows us to take \( p_U \) as the uniform distribution over the disk (i.e., \( p_U = 1/\pi \)) with no learnable parameters. Consequently, this model places no probability mass outside the disk, needing no truncation.

Despite their better expressiveness in general, the two off-the-shelf NF models introduced above remain (costly) monolithic transformations, sacrificing the interpretability inherent to mixture models, a trade-off we will examine next in Section \ref{sec:eval}.

\section{Evaluation Results}\label{sec:eval}

In this section, we evaluate the performance of MDNs (i.e., mixing Gaussian, truncated Gaussian, beta-von-Mises, and M\"obius distributions) and compare them against NF models (i.e., RNVP, truncated RNVP, and NSF). For the truncated versions of the Gaussian mixture and RNVP models, we adopt a simple, approximate renormalization method; i.e., density values are evaluated only on \(\mathcal{S}\) and renormalized by a grid-based Riemann sum, rather than using exact truncation normalizers, which are computationally expensive or even prohibitive (e.g., for RNVP).

Without loss of generality, the results shown here are the outcomes of 10 independent training runs (using different seeds), evaluated under stratified 5-fold cross-validation on 30 settings of $\lambda=(d, d', r_c, \theta_c)$, with $d=0.2$ and $\theta_c=\pi$ held fixed and $(d', r_c)$ ranging over the $3\times 10$ grid, $\{0.1, 0.2, 0.3\}\times\{0.0, 0.1, \ldots, 0.9\}$. Each fold trains on 24 examples and evaluates generalization on 6 held-out settings that are never used for training or model selection. Our evaluation spans multiple dimensions: quantitative accuracy on the training and unseen held-out sets in Section \ref{sec:eval_quant} (Tables \ref{tab:g_vs_m} and \ref{tab:nf}, Figures \ref{fig:train_perf_mdn}, \ref{fig:pred_perf_mdn}, \ref{fig:train_perf_nf}, and \ref{fig:pred_perf_nf}), visual interpretability of the generated distributions in Section \ref{sec:eval_interp} (Figures \ref{fig:g_m} and \ref{fig:nf}), a case study of prediction performance under data scarcity in Section \ref{sec:prediction} (Figure \ref{fig:mob_nsf}), analyses of training dynamics in Section \ref{sec:training} (Figures \ref{fig:train_loss_mdn} and \ref{fig:train_loss_nf}), and finally inference computation performance evaluation in Section \ref{sec:inference}.

\subsection{Quantitative Performance and Generalizability}\label{sec:eval_quant}

\begin{figure*}[!t]
    \centering
    
    \makeatletter\def\@captype{table}\makeatother
    \setlength{\tabcolsep}{0.5em}
    {\renewcommand{\arraystretch}{0.75}
    \footnotesize
    \begin{tabular}{|c|c|c|c|c|c|c|c|c|c|c|c|}
         \cline{2-12}
         \multicolumn{1}{c|}{} & $K$ & 1 & 2 & 3 & 4 & 5 & 6 & 7 & 8 & 9 & 10 \\
         \hhline{|=|=|=|=|=|=|=|=|=|=|=|=|}
         \multirow{5}{*}[0.1em]{\rotatebox[origin=c]{90}{Gaussian}} 
         & Params. & 238 & 828 & 1,770 & 3,064 & 4,710 & 6,708 & 9,058 & 11,760 & 14,814 & 18,220 \\
         \cline{2-12} \\[-1.7em]
         & MSE & \makecell[r]{$0.03036$\\[-0.8em] $\pm 0.00000$}
               & \makecell[r]{$0.00938$\\[-0.8em] $\pm 0.00000$}
               & \makecell[r]{$0.01167$\\[-0.8em] $\pm 0.00082$}
               & \makecell[r]{$0.00785$\\[-0.8em] $\pm 0.00000$}
               & \makecell[r]{$0.00702$\\[-0.8em] $\pm 0.00034$}
               & \makecell[r]{$0.00428$\\[-0.8em] $\pm 0.00077$}
               & \makecell[r]{$0.00195$\\[-0.8em] $\pm 0.00034$}
               & \makecell[r]{$0.00110$\\[-0.8em] $\pm 0.00018$}
               & \makecell[r]{$0.00066$\\[-0.8em] $\pm 0.00004$}
               & \makecell[r]{$0.00059$\\[-0.8em] $\pm 0.00005$} \\[-0.2em]
         \cline{2-12} \\[-1.7em]
         & GKL
               & \makecell[r]{$0.05132$\\[-0.8em] $\pm 0.00000$}
               & \makecell[r]{$0.02578$\\[-0.8em] $\pm 0.00000$}
               & \makecell[r]{$0.02261$\\[-0.8em] $\pm 0.00052$}
               & \makecell[r]{$0.01268$\\[-0.8em] $\pm 0.00000$}
               & \makecell[r]{$0.01086$\\[-0.8em] $\pm 0.00008$}
               & \makecell[r]{$0.00859$\\[-0.8em] $\pm 0.00029$}
               & \makecell[r]{$0.00639$\\[-0.8em] $\pm 0.00013$}
               & \makecell[r]{$0.00529$\\[-0.8em] $\pm 0.00010$}
               & \makecell[r]{$0.00458$\\[-0.8em] $\pm 0.00003$}
               & \makecell[r]{$0.00402$\\[-0.8em] $\pm 0.00003$} \\[-0.2em]
         \hhline{|=|=|=|=|=|=|=|=|=|=|=|=|}
         \multirow{5}{*}[0.1em]{\rotatebox[origin=c]{90}{\shortstack[c]{Gaussian\\Truncated}}} 
         & Params. & 238 & 828 & 1,770 & 3,064 & 4,710 & 6,708 & 9,058 & 11,760 & 14,814 & 18,220 \\
         \cline{2-12} \\[-1.7em]
         & MSE & \makecell[r]{$0.02971$\\[-0.8em] $\pm 0.00000$}
               & \makecell[r]{$0.00855$\\[-0.8em] $\pm 0.00000$}
               & \makecell[r]{$0.00865$\\[-0.8em] $\pm 0.00065$}
               & \makecell[r]{$0.00755$\\[-0.8em] $\pm 0.00000$}
               & \makecell[r]{$0.00660$\\[-0.8em] $\pm 0.00029$}
               & \makecell[r]{$0.00318$\\[-0.8em] $\pm 0.00048$}
               & \makecell[r]{$0.00174$\\[-0.8em] $\pm 0.00021$}
               & \makecell[r]{$0.00082$\\[-0.8em] $\pm 0.00017$}
               & \makecell[r]{$0.00060$\\[-0.8em] $\pm 0.00005$}
               & \makecell[r]{$0.00055$\\[-0.8em] $\pm 0.00005$} \\[-0.2em]
         \cline{2-12} \\[-1.7em]
         & GKL
               & \makecell[r]{$0.04793$\\[-0.8em] $\pm 0.00000$}
               & \makecell[r]{$0.02239$\\[-0.8em] $\pm 0.00000$}
               & \makecell[r]{$0.01912$\\[-0.8em] $\pm 0.00048$}
               & \makecell[r]{$0.01199$\\[-0.8em] $\pm 0.00000$}
               & \makecell[r]{$0.01029$\\[-0.8em] $\pm 0.00009$}
               & \makecell[r]{$0.00784$\\[-0.8em] $\pm 0.00019$}
               & \makecell[r]{$0.00612$\\[-0.8em] $\pm 0.00009$}
               & \makecell[r]{$0.00498$\\[-0.8em] $\pm 0.00008$}
               & \makecell[r]{$0.00442$\\[-0.8em] $\pm 0.00003$}
               & \makecell[r]{$0.00390$\\[-0.8em] $\pm 0.00005$} \\[-0.2em]
         \hhline{|=|=|=|=|=|=|=|=|=|=|=|=|}
         \multirow{5}{*}[0.1em]{\rotatebox[origin=c]{90}{\shortstack[c]{Beta $\times$\\von Mises}}} 
         & Params. & 229 & 794 & 1,695 & 2,932 & 4,505 & 6,414 & 8,659 & 11,240 & 14,157 & 17,410 \\
         \cline{2-12} \\[-1.7em]
         & MSE & \makecell[r]{$0.02468$\\[-0.8em] $\pm 0.00002$}
               & \makecell[r]{$0.01357$\\[-0.8em] $\pm 0.00022$}
               & \makecell[r]{$0.01046$\\[-0.8em] $\pm 0.00006$}
               & \makecell[r]{$0.00893$\\[-0.8em] $\pm 0.00003$}
               & \makecell[r]{$0.00785$\\[-0.8em] $\pm 0.00003$}
               & \makecell[r]{$0.00706$\\[-0.8em] $\pm 0.00005$}
               & \makecell[r]{$0.00639$\\[-0.8em] $\pm 0.00002$}
               & \makecell[r]{$0.00598$\\[-0.8em] $\pm 0.00003$}
               & \makecell[r]{$0.00561$\\[-0.8em] $\pm 0.00003$}
               & \makecell[r]{$0.00535$\\[-0.8em] $\pm 0.00003$} \\[-0.2em]
         \cline{2-12} \\[-1.7em]
         & GKL
               & \makecell[r]{$0.02721$\\[-0.8em] $\pm 0.00004$}
               & \makecell[r]{$0.01086$\\[-0.8em] $\pm 0.00027$}
               & \makecell[r]{$0.00741$\\[-0.8em] $\pm 0.00007$}
               & \makecell[r]{$0.00584$\\[-0.8em] $\pm 0.00003$}
               & \makecell[r]{$0.00489$\\[-0.8em] $\pm 0.00003$}
               & \makecell[r]{$0.00426$\\[-0.8em] $\pm 0.00003$}
               & \makecell[r]{$0.00378$\\[-0.8em] $\pm 0.00002$}
               & \makecell[r]{$0.00350$\\[-0.8em] $\pm 0.00002$}
               & \makecell[r]{$0.00326$\\[-0.8em] $\pm 0.00001$}
               & \makecell[r]{$0.00309$\\[-0.8em] $\pm 0.00001$} \\[-0.2em]
         \hhline{|=|=|=|=|=|=|=|=|=|=|=|=|}
         \multirow{5}{*}[0.1em]{\rotatebox[origin=c]{90}{M\"obius}} 
         & Params. & 229 & 794 & 1,695 & 2,932 & 4,505 & 6,414 & 8,659 & 11,240 & 14,157 & 17,410 \\
         \cline{2-12} \\[-1.7em]
         & MSE & \makecell[r]{$0.02653$\\[-0.8em] $\pm 0.00003$}
               & \makecell[r]{$0.00730$\\[-0.8em] $\pm 0.00052$}
               & \makecell[r]{$0.00137$\\[-0.8em] $\pm 0.00001$}
               & \makecell[r]{$0.00137$\\[-0.8em] $\pm 0.00004$}
               & \makecell[r]{$0.00126$\\[-0.8em] $\pm 0.00006$}
               & \makecell[r]{$0.00093$\\[-0.8em] $\pm 0.00006$}
               & \makecell[r]{$0.00062$\\[-0.8em] $\pm 0.00014$}
               & \makecell[r]{$0.00037$\\[-0.8em] $\pm 0.00005$}
               & \makecell[r]{$0.00032$\\[-0.8em] $\pm 0.00006$}
               & \makecell[r]{$0.00022$\\[-0.8em] $\pm 0.00003$} \\[-0.2em]
         \cline{2-12} \\[-1.7em]
         & GKL
               & \makecell[r]{$0.03342$\\[-0.8em] $\pm 0.00008$}
               & \makecell[r]{$0.01186$\\[-0.8em] $\pm 0.00062$}
               & \makecell[r]{$0.00154$\\[-0.8em] $\pm 0.00001$}
               & \makecell[r]{$0.00185$\\[-0.8em] $\pm 0.00020$}
               & \makecell[r]{$0.00176$\\[-0.8em] $\pm 0.00017$}
               & \makecell[r]{$0.00115$\\[-0.8em] $\pm 0.00011$}
               & \makecell[r]{$0.00083$\\[-0.8em] $\pm 0.00010$}
               & \makecell[r]{$0.00060$\\[-0.8em] $\pm 0.00005$}
               & \makecell[r]{$0.00054$\\[-0.8em] $\pm 0.00005$}
               & \makecell[r]{$0.00045$\\[-0.8em] $\pm 0.00003$} \\[-0.2em]
         \hhline{|=|=|=|=|=|=|=|=|=|=|=|=|}
    \end{tabular}
    }
    \caption{Training set (with 24 training points per fold) performance statistics of the MDN models over 10 runs, including 95\% confidence intervals.}
    \label{tab:g_vs_m}
    
    \vspace{2em}
    
    \makeatletter\def\@captype{figure}\makeatother
    \subfigure[]{
    \label{fig:train_perf_mdn:a}
    \includegraphics[width=3.5in]{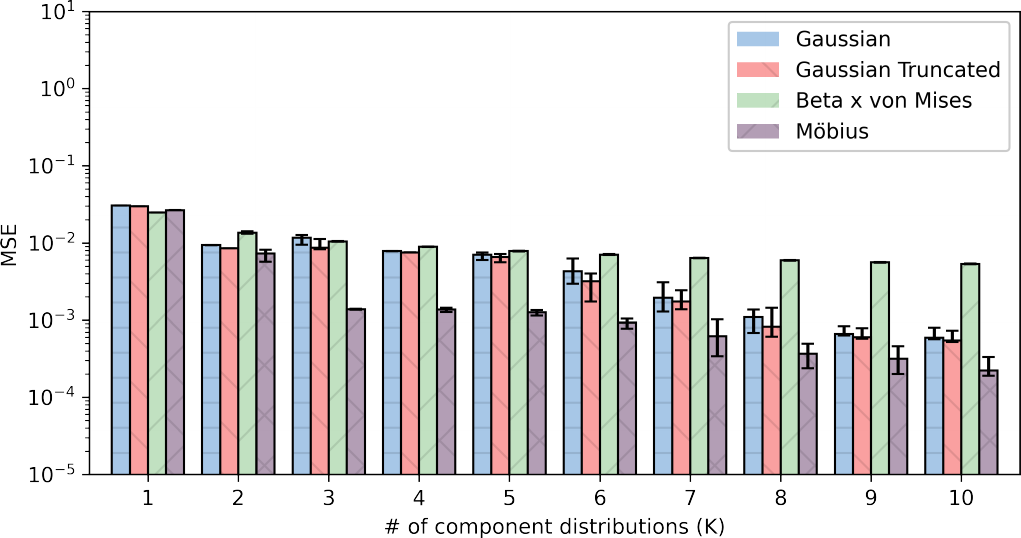}}
    \subfigure[]{
    \label{fig:train_perf_mdn:b}
    \includegraphics[width=3.465in]{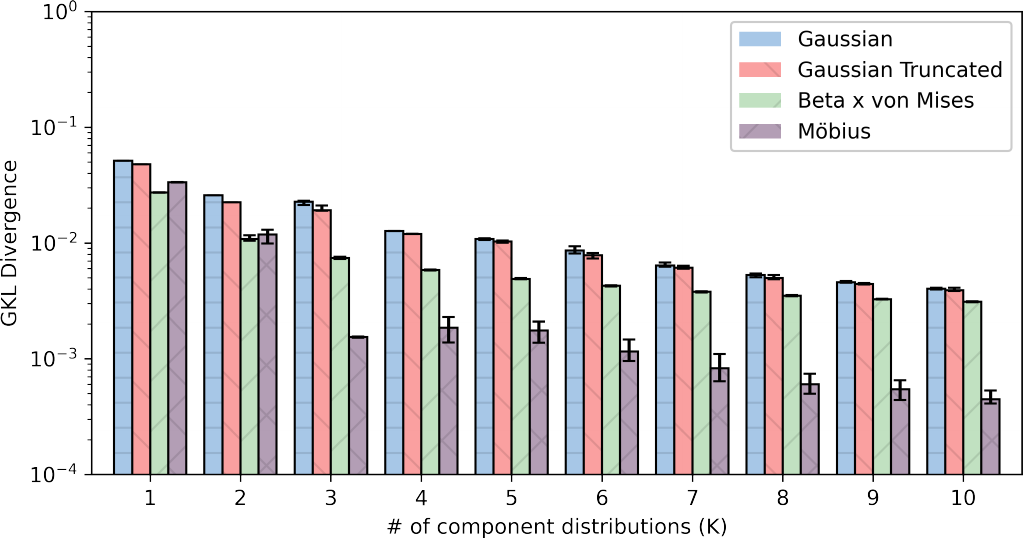}}
    \caption{Training set (with 24 training points per fold) performance of the MDN models over 10 runs, with error bars from minimum to maximum values.}
    \label{fig:train_perf_mdn}
    
    \vspace{2em}
    
    \subfigure[]{
    \label{fig:pred_perf_mdn:a}
    \includegraphics[width=3.5in]{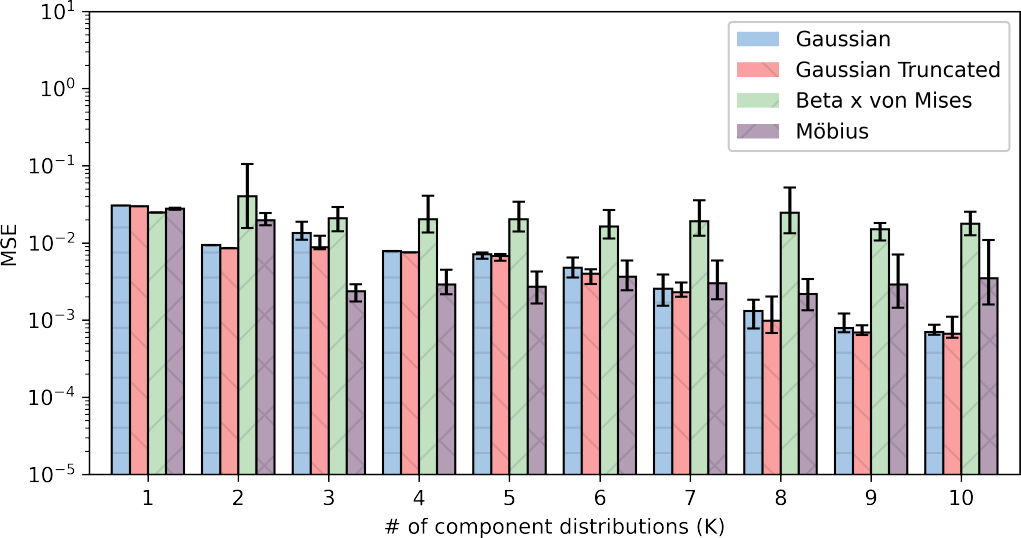}}
    \subfigure[]{
    \label{fig:pred_perf_mdn:b}
    \includegraphics[width=3.465in]{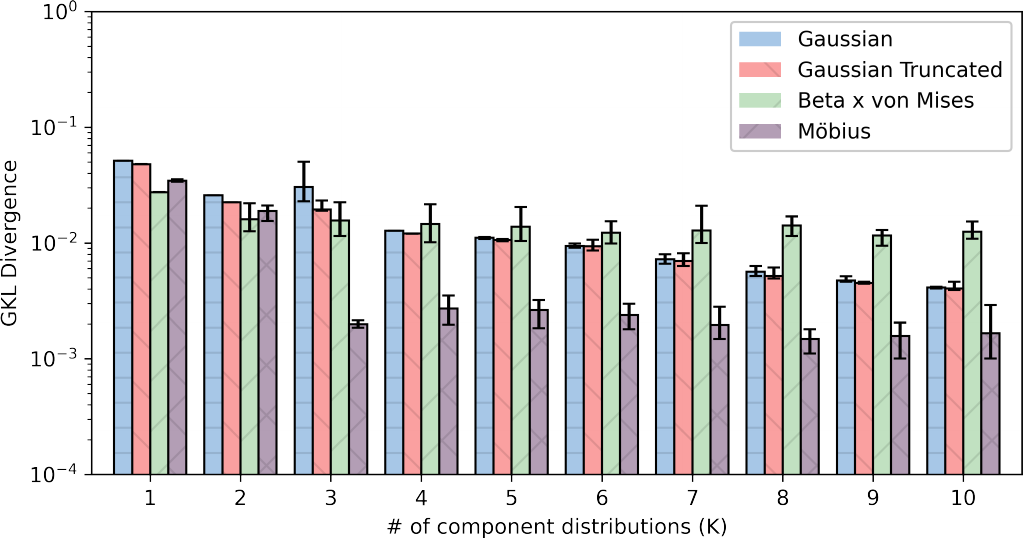}}
    \caption{Prediction performance (with 6 prediction points per fold) of the MDN models over 10 runs, with error bars from minimum to maximum values.}
    \label{fig:pred_perf_mdn}
\end{figure*}

\begin{figure*}[!t]
      \centering
      
      \makeatletter\def\@captype{table}\makeatother
      \setlength{\tabcolsep}{0.5em}
      {\renewcommand{\arraystretch}{0.75}
      \footnotesize
      \begin{tabular}{|c|c|c|c|c|c|c|c|c|c|c|c|}
           \cline{2-12}
           \multicolumn{1}{c|}{} & \(L\) & 1 & 2 & 3 & 4 & 5 & 6 & 7 & 8 & 9 & 10 \\
           \hhline{|=|=|=|=|=|=|=|=|=|=|=|=|}
           \multirow{5}{*}[0.1em]{\rotatebox[origin=c]{90}{RNVP}} 
           & Params. & 9,056 & 17,955 & 26,854 & 35,753 & 44,652 & 53,551 & 62,450 & 71,349 & 80,248 & 89,147 \\
           \cline{2-12} \\[-1.7em]
           & MSE & \makecell[r]{\(0.02472\)\\[-0.8em] \(\pm 0.00035\)}
                 & \makecell[r]{\(0.00563\)\\[-0.8em] \(\pm 0.00095\)}
                 & \makecell[r]{\(0.00314\)\\[-0.8em] \(\pm 0.00031\)}
                 & \makecell[r]{\(0.00361\)\\[-0.8em] \(\pm 0.00047\)}
                 & \makecell[r]{\(0.00354\)\\[-0.8em] \(\pm 0.00053\)}
                 & \makecell[r]{\(0.00153\)\\[-0.8em] \(\pm 0.00035\)}
                 & \makecell[r]{\(0.00209\)\\[-0.8em] \(\pm 0.00027\)}
                 & \makecell[r]{\(0.00232\)\\[-0.8em] \(\pm 0.00054\)}
                 & \makecell[r]{\(0.00253\)\\[-0.8em] \(\pm 0.00037\)}
                 & \makecell[r]{\(0.00164\)\\[-0.8em] \(\pm 0.00025\)} \\[-0.2em]
           \cline{2-12} \\[-1.7em]
           & GKL
                 & \makecell[r]{\(0.02536\)\\[-0.8em] \(\pm 0.00030\)}
                 & \makecell[r]{\(0.00463\)\\[-0.8em] \(\pm 0.00098\)}
                 & \makecell[r]{\(0.00301\)\\[-0.8em] \(\pm 0.00027\)}
                 & \makecell[r]{\(0.00349\)\\[-0.8em] \(\pm 0.00037\)}
                 & \makecell[r]{\(0.00352\)\\[-0.8em] \(\pm 0.00045\)}
                 & \makecell[r]{\(0.00181\)\\[-0.8em] \(\pm 0.00027\)}
                 & \makecell[r]{\(0.00207\)\\[-0.8em] \(\pm 0.00021\)}
                 & \makecell[r]{\(0.00232\)\\[-0.8em] \(\pm 0.00045\)}
                 & \makecell[r]{\(0.00251\)\\[-0.8em] \(\pm 0.00029\)}
                 & \makecell[r]{\(0.00190\)\\[-0.8em] \(\pm 0.00020\)} \\[-0.2em]
           \hhline{|=|=|=|=|=|=|=|=|=|=|=|=|}
           \multirow{5}{*}[0.1em]{\rotatebox[origin=c]{90}{\shortstack[c]{RNVP\\Truncated}}} 
           & Params. & 9,056 & 17,955 & 26,854 & 35,753 & 44,652 & 53,551 & 62,450 & 71,349 & 80,248 & 89,147 \\
           \cline{2-12} \\[-1.7em]
           & MSE & \makecell[r]{\(0.02193\)\\[-0.8em] \(\pm 0.00038\)}
                 & \makecell[r]{\(0.00284\)\\[-0.8em] \(\pm 0.00085\)}
                 & \makecell[r]{\(0.00319\)\\[-0.8em] \(\pm 0.00099\)}
                 & \makecell[r]{\(0.00331\)\\[-0.8em] \(\pm 0.00126\)}
                 & \makecell[r]{\(0.00292\)\\[-0.8em] \(\pm 0.00085\)}
                 & \makecell[r]{\(0.00249\)\\[-0.8em] \(\pm 0.00048\)}
                 & \makecell[r]{\(0.00274\)\\[-0.8em] \(\pm 0.00051\)}
                 & \makecell[r]{\(0.00242\)\\[-0.8em] \(\pm 0.00047\)}
                 & \makecell[r]{\(0.00310\)\\[-0.8em] \(\pm 0.00089\)}
                 & \makecell[r]{\(0.00284\)\\[-0.8em] \(\pm 0.00099\)} \\[-0.2em]
           \cline{2-12} \\[-1.7em]
           & GKL
                 & \makecell[r]{\(0.02308\)\\[-0.8em] \(\pm 0.00052\)}
                 & \makecell[r]{\(0.00346\)\\[-0.8em] \(\pm 0.00090\)}
                 & \makecell[r]{\(0.00341\)\\[-0.8em] \(\pm 0.00097\)}
                 & \makecell[r]{\(0.00338\)\\[-0.8em] \(\pm 0.00117\)}
                 & \makecell[r]{\(0.00311\)\\[-0.8em] \(\pm 0.00084\)}
                 & \makecell[r]{\(0.00255\)\\[-0.8em] \(\pm 0.00038\)}
                 & \makecell[r]{\(0.00269\)\\[-0.8em] \(\pm 0.00041\)}
                 & \makecell[r]{\(0.00252\)\\[-0.8em] \(\pm 0.00042\)}
                 & \makecell[r]{\(0.00306\)\\[-0.8em] \(\pm 0.00093\)}
                 & \makecell[r]{\(0.00286\)\\[-0.8em] \(\pm 0.00103\)} \\[-0.2em]
           \hhline{|=|=|=|=|=|=|=|=|=|=|=|=|}
           \multirow{5}{*}[0.1em]{\rotatebox[origin=c]{90}{NSF}} 
           & Params. & 10,199 & 20,398 & 30,597 & 40,796 & 50,995 & 61,194 & 71,393 & 81,592 & 91,791 & 101,990 \\
           \cline{2-12} \\[-1.7em]
           & MSE & \makecell[r]{\(0.04269\)\\[-0.8em] \(\pm 0.01689\)}
                 & \makecell[r]{\(0.00008\)\\[-0.8em] \(\pm 0.00001\)}
                 & \makecell[r]{\(0.00006\)\\[-0.8em] \(\pm 0.00000\)}
                 & \makecell[r]{\(0.00005\)\\[-0.8em] \(\pm 0.00000\)}
                 & \makecell[r]{\(0.00008\)\\[-0.8em] \(\pm 0.00001\)}
                 & \makecell[r]{\(0.00004\)\\[-0.8em] \(\pm 0.00000\)}
                 & \makecell[r]{\(0.00005\)\\[-0.8em] \(\pm 0.00000\)}
                 & \makecell[r]{\(0.00006\)\\[-0.8em] \(\pm 0.00000\)}
                 & \makecell[r]{\(0.00006\)\\[-0.8em] \(\pm 0.00000\)}
                 & \makecell[r]{\(0.00005\)\\[-0.8em] \(\pm 0.00000\)} \\[-0.2em]
           \cline{2-12} \\[-1.7em]
           & GKL
                 & \makecell[r]{\(0.13710\)\\[-0.8em] \(\pm 0.03490\)}
                 & \makecell[r]{\(0.00088\)\\[-0.8em] \(\pm 0.00010\)}
                 & \makecell[r]{\(0.00105\)\\[-0.8em] \(\pm 0.00001\)}
                 & \makecell[r]{\(0.00026\)\\[-0.8em] \(\pm 0.00008\)}
                 & \makecell[r]{\(0.00033\)\\[-0.8em] \(\pm 0.00008\)}
                 & \makecell[r]{\(0.00068\)\\[-0.8em] \(\pm 0.00001\)}
                 & \makecell[r]{\(0.00071\)\\[-0.8em] \(\pm 0.00002\)}
                 & \makecell[r]{\(0.00077\)\\[-0.8em] \(\pm 0.00002\)}
                 & \makecell[r]{\(0.00078\)\\[-0.8em] \(\pm 0.00003\)}
                 & \makecell[r]{\(0.00074\)\\[-0.8em] \(\pm 0.00001\)} \\[-0.2em]
           \hhline{|=|=|=|=|=|=|=|=|=|=|=|=|}
      \end{tabular}
      }
      \caption{Training set (with 24 training points per fold) performance statistics of the NF models over 10 runs, including 95\% confidence intervals.}
      \label{tab:nf}
      
      \vspace{2em}
      
      \makeatletter\def\@captype{figure}\makeatother
      \subfigure[]{
      \label{fig:train_perf_nf:a}
      \includegraphics[width=3.5in]{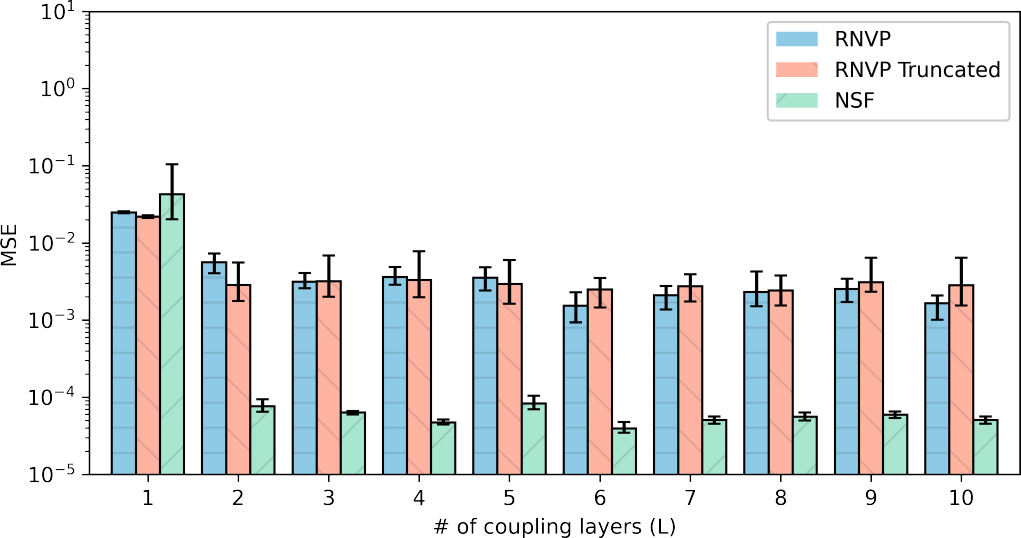}}
      \subfigure[]{
      \label{fig:train_perf_nf:b}
      \includegraphics[width=3.465in]{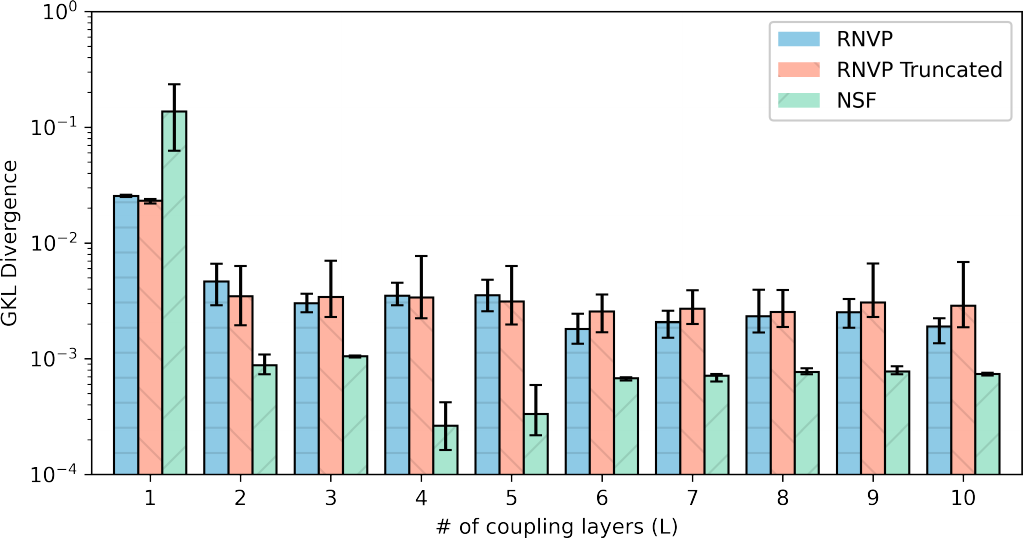}}
      \caption{Training set (with 24 training points per fold) performance of the NF models over 10 runs, with error bars from minimum to maximum values.}
      \label{fig:train_perf_nf}
      
      \vspace{2em}
      
      \subfigure[]{
      \label{fig:pred_perf_nf:a}
      \includegraphics[width=3.5in]{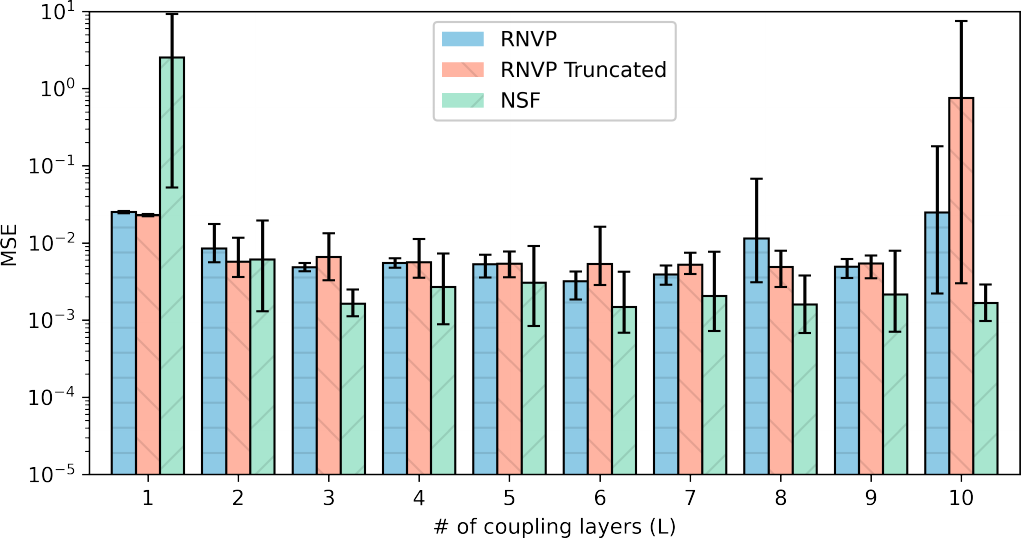}}
      \subfigure[]{
      \label{fig:pred_perf_nf:b}
      \includegraphics[width=3.465in]{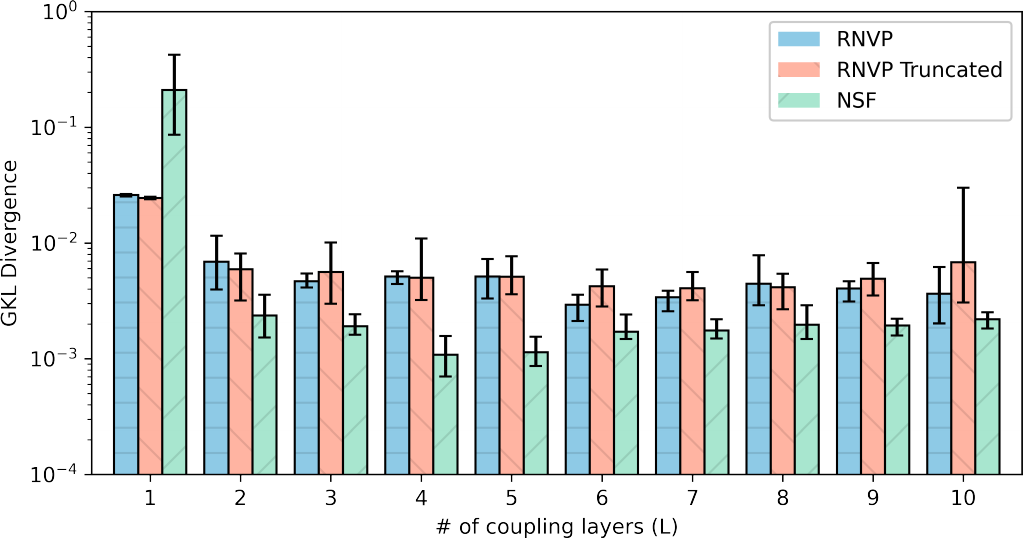}}
      \caption{Prediction performance (with 6 prediction points per fold) of the NF models over 10 runs, with error bars from minimum to maximum values.}
      \label{fig:pred_perf_nf}
  \end{figure*}

A summary of the quantitative training results is shown in Table \ref{tab:g_vs_m} for the MDNs and Table \ref{tab:nf} for the NFs. The columns indicate the complexity of the models: the number of mixture components, $K$, for MDNs, and the number of affine/spline coupling layers, $L$, for NFs, respectively. Notably, the row \textit{Params.} lists the total number of learnable parameters (i.e., the weights and biases of the underlying neural networks) required by each model. %

The results are shown in terms of MSE (measuring pointwise accuracy) and GKL divergence (for distributional discrepancy), evaluated over the disk support $\mathcal{S}$, accompanied by 95\% confidence intervals (Tables \ref{tab:g_vs_m} and \ref{tab:nf}) or min-max error bars (Figures \ref{fig:train_perf_mdn}, \ref{fig:pred_perf_mdn}, \ref{fig:train_perf_nf}, and \ref{fig:pred_perf_nf}). Looking at the MDN models through Table \ref{tab:g_vs_m} and Figure \ref{fig:train_perf_mdn}, given that the training data set is the same, the benefit of using the M\"obius kernel is obvious under both the MSE and GKL-divergence metrics. For all \(3 \leq K \leq 10\), the M\"obius model excels at fitting target distributions in the training sets, with its MSE and GKL divergence remaining superior to those of the other three mixture models across the ten independent training runs (illustrated by the error bars in Figure \ref{fig:train_perf_mdn}). In particular, the markedly improved GKL score suggests a better matching of the overall shape of the distribution using M\"obius.

Regarding the NF models (Table \ref{tab:nf} and Figure \ref{fig:train_perf_nf}), we find that they can achieve competitive and even superior (for NSF with $L=4$ or $5$ particularly) MSE and GKL scores relative to the M\"obius MDN once a  number of $L\geq 2$ coupling layers is used. This verifies that the NFs have the expressiveness to finely capture the target node density, but with the caveat that the numbers of learnable parameters required for them to gain enough capacity are on the order of \(\mathcal{O}(10^4)\) or even \(\mathcal{O}(10^5)\) (e.g., up to 101,990 for NSF with $L=10$), whereas the M\"obius mixture can reach a strong fit with a few thousand parameters (e.g., 1,695 at \(K=3\) and under 19,000 even for $K=10$). This highlights a trade-off between NFs and MDNs, i.e., the former rely on architectural sophistication and a heavier parameterization, while the latter can attain comparable accuracy far more frugally by embedding geometrical inductive bias in the mixture kernels.

A demonstration that the training has been effective can be witnessed in the prediction results shown in Figure \ref{fig:pred_perf_mdn} (for MDNs) and Figure \ref{fig:pred_perf_nf} (for NFs). For both MDNs and NFs, the prediction results generally align well with their respective training performances, indicating no drastic overfitting.

For the MDNs, this alignment demonstrates that they can well generalize reliably from the small training set of 24 points (per fold), which is practically important for applications involving expensive simulations. Again, the M\"obius model maintains a consistently lower GKL divergence for \(3\leq K\leq 9\) compared to the other three mixture models, although the performance gap narrows relative to the Gaussian models. Due to its radial-angular independence, the beta-von-Mises mixture shows the worst generalizability among the MDNs, with GKL divergence (and MSE) being the highest for most $K$ settings. For the NF models, the prediction results broadly preserve the training trends, albeit with a larger train-to-prediction gap in MSE for NSF for all $L$ settings. Both NSF and the two RNVP variants improve substantially from \(L=2\), but beyond this initial improvement, increasing \(L\) yields no consistent generalization gain, while the wider min-max error bars for RNVPs at certain larger values of \(L\) (e.g., $L=10$) indicate occasional unstable runs.

\subsection{Interpreting the Mixture Results}\label{sec:eval_interp}

The differences among the models, particularly that MDNs (e.g., the M\"obius mixture) remain interpretable relative to the NFs, can be explained by the decomposability of the former, which enable component-level inspection of mobility effects (i.e., charger attraction versus RWP), and by the structural bias of each type of kernel.

\begin{figure*}[!t]
 \centering
 \subfigure[]{
 \label{fig:g_m:a}
 \includegraphics[width=5.4cm]{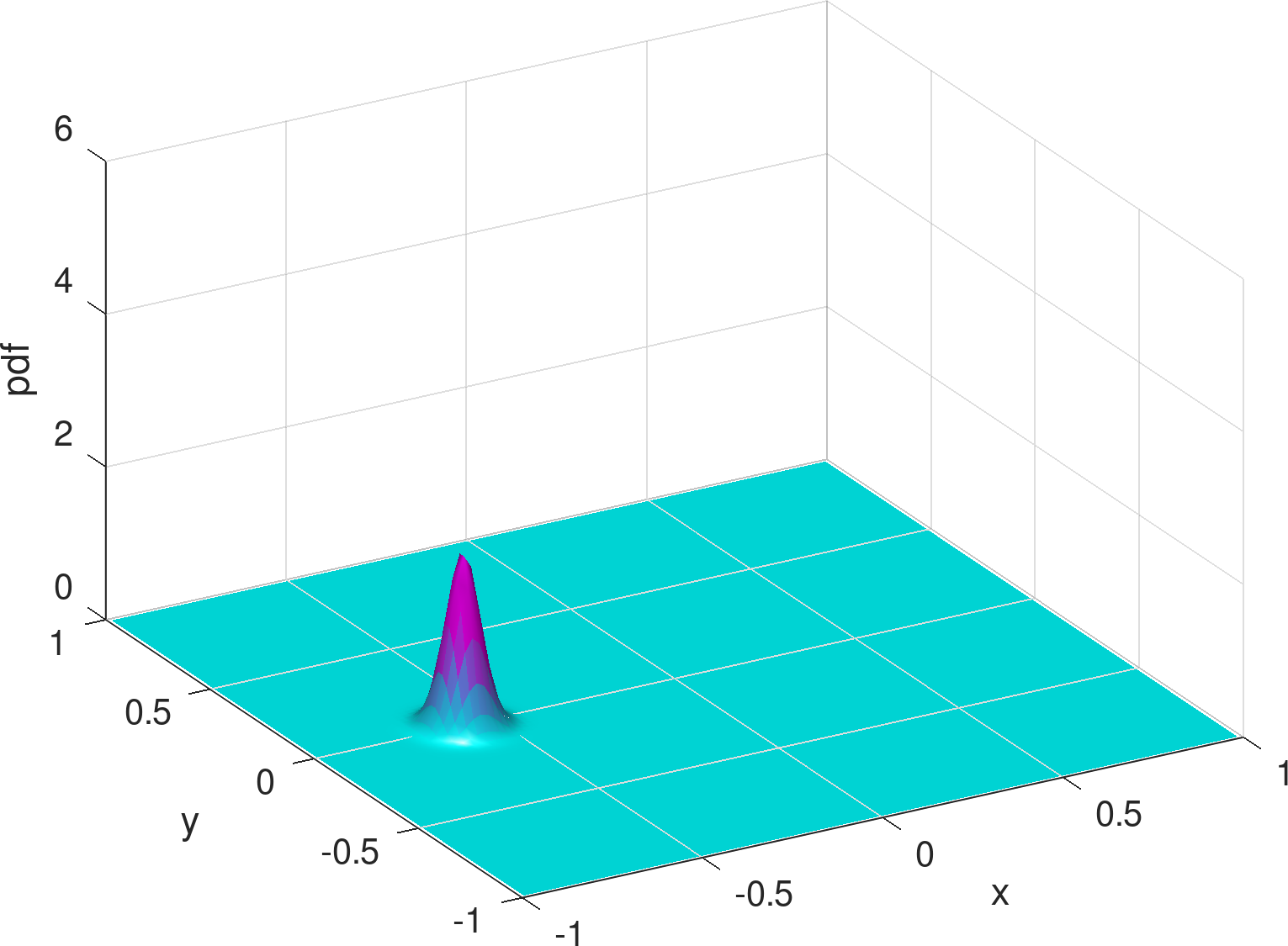}}
 \subfigure[]{
 \label{fig:g_m:b}
 \includegraphics[width=5.4cm]{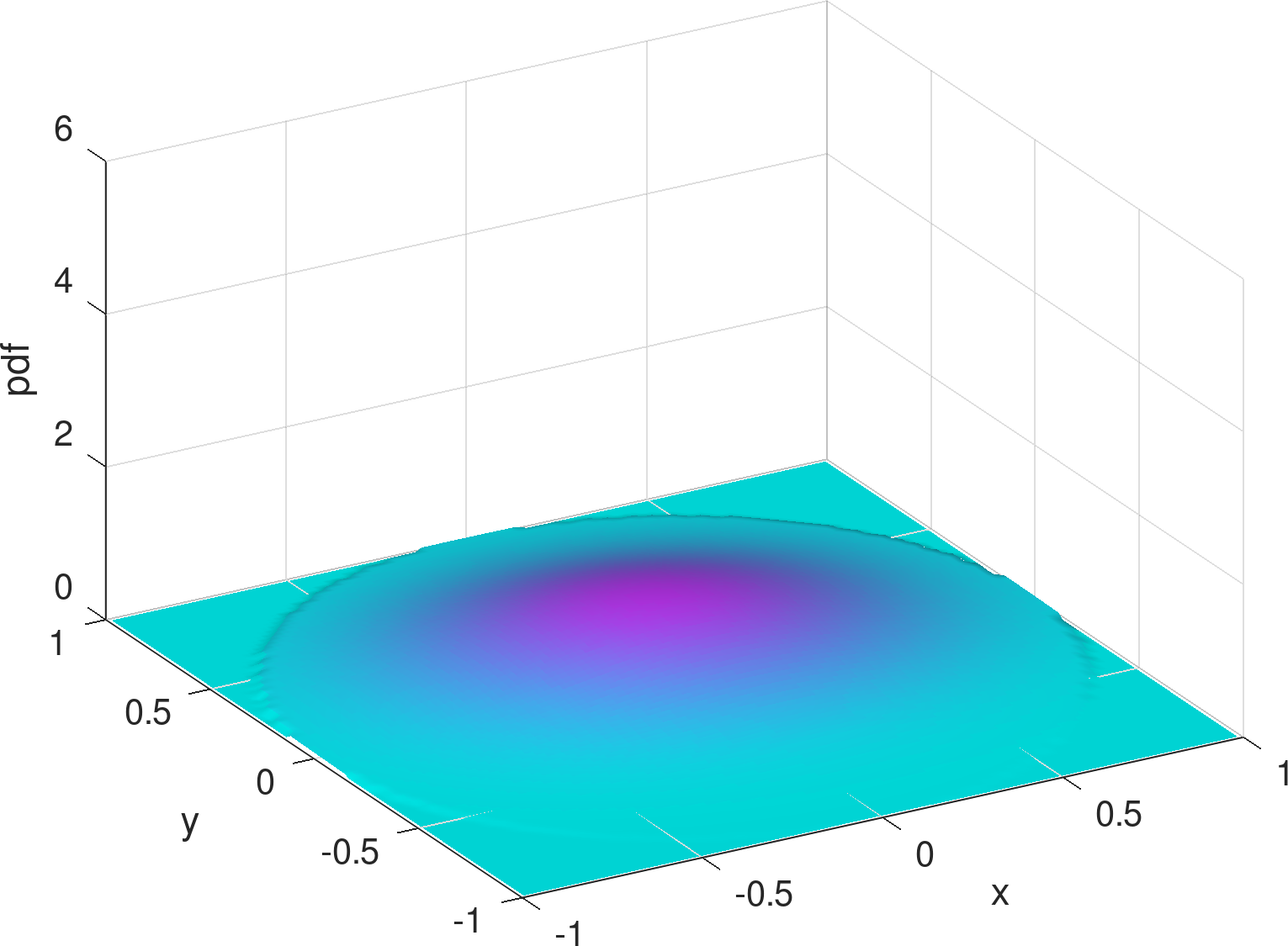}}
 \subfigure[]{
 \label{fig:g_m:c}
 \includegraphics[width=5.4cm]{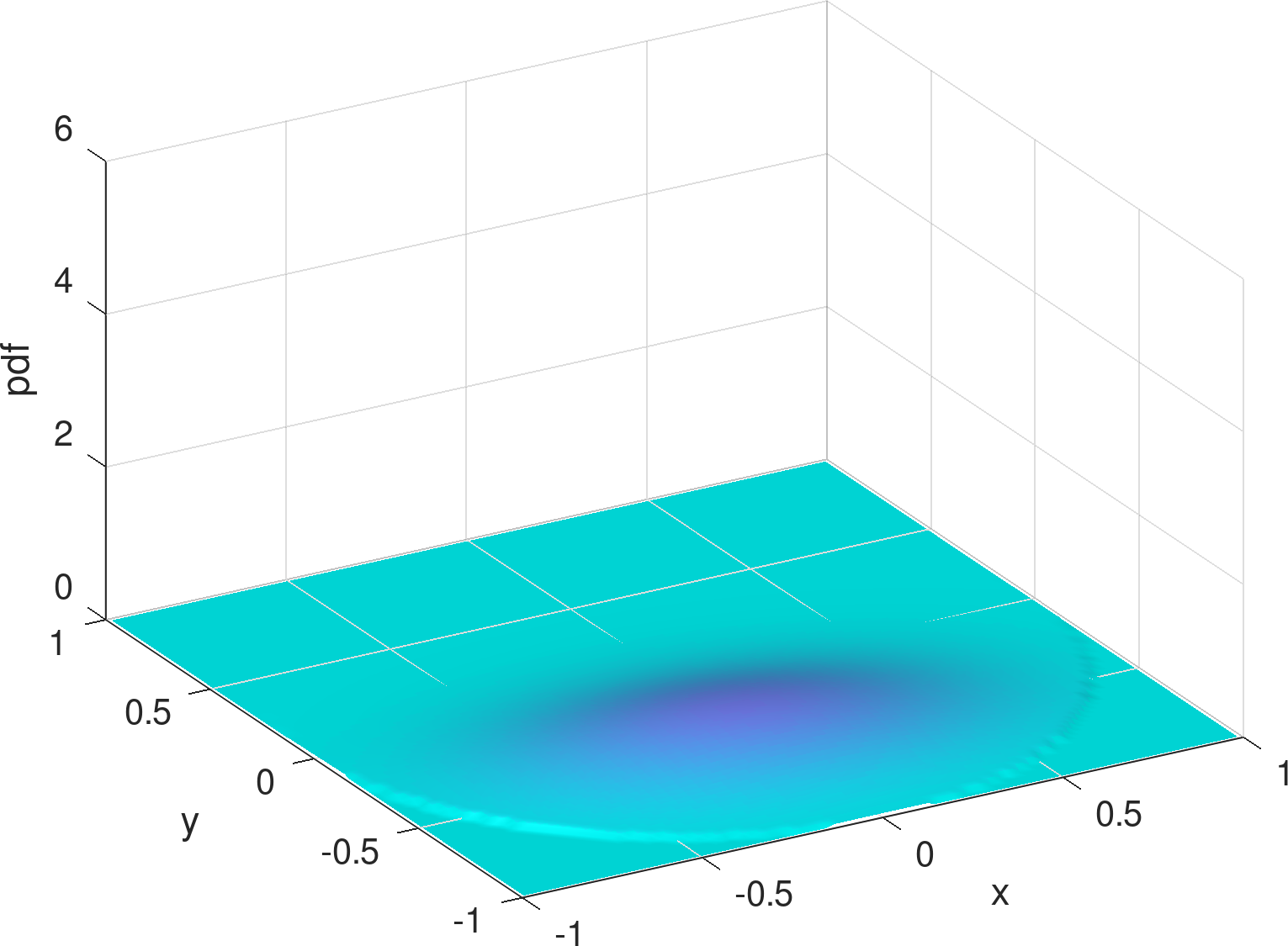}}
 \subfigure[]{
 \label{fig:g_m:d}
 \includegraphics[width=5.4cm]{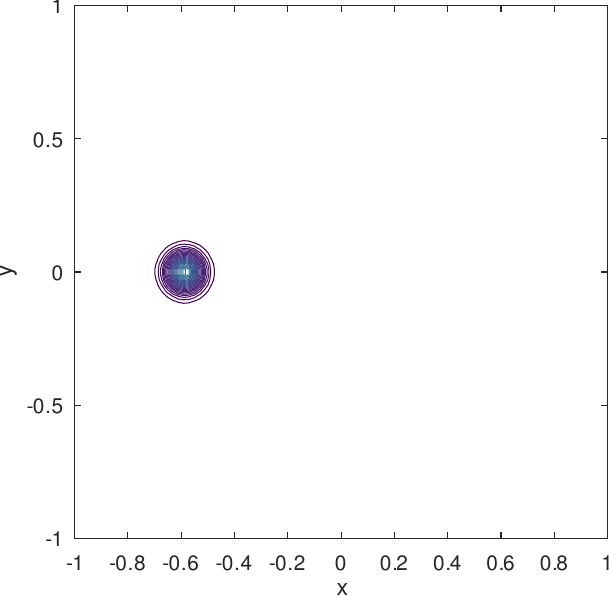}}
 \subfigure[]{
 \label{fig:g_m:e}
 \includegraphics[width=5.4cm]{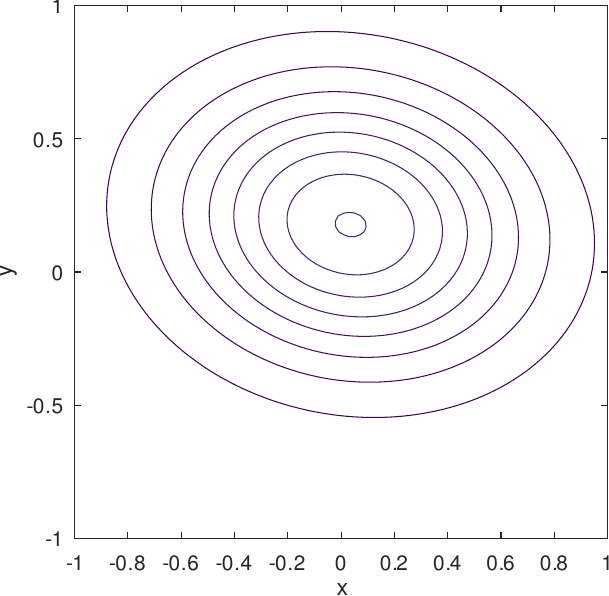}}
 \subfigure[]{
 \label{fig:g_m:f}
 \includegraphics[width=5.4cm]{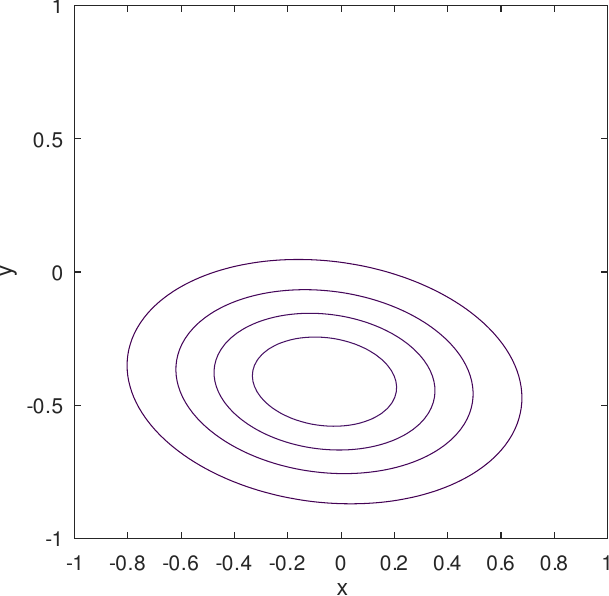}}
 \subfigure[]{
 \label{fig:g_m:h}
 \includegraphics[width=5.4cm]{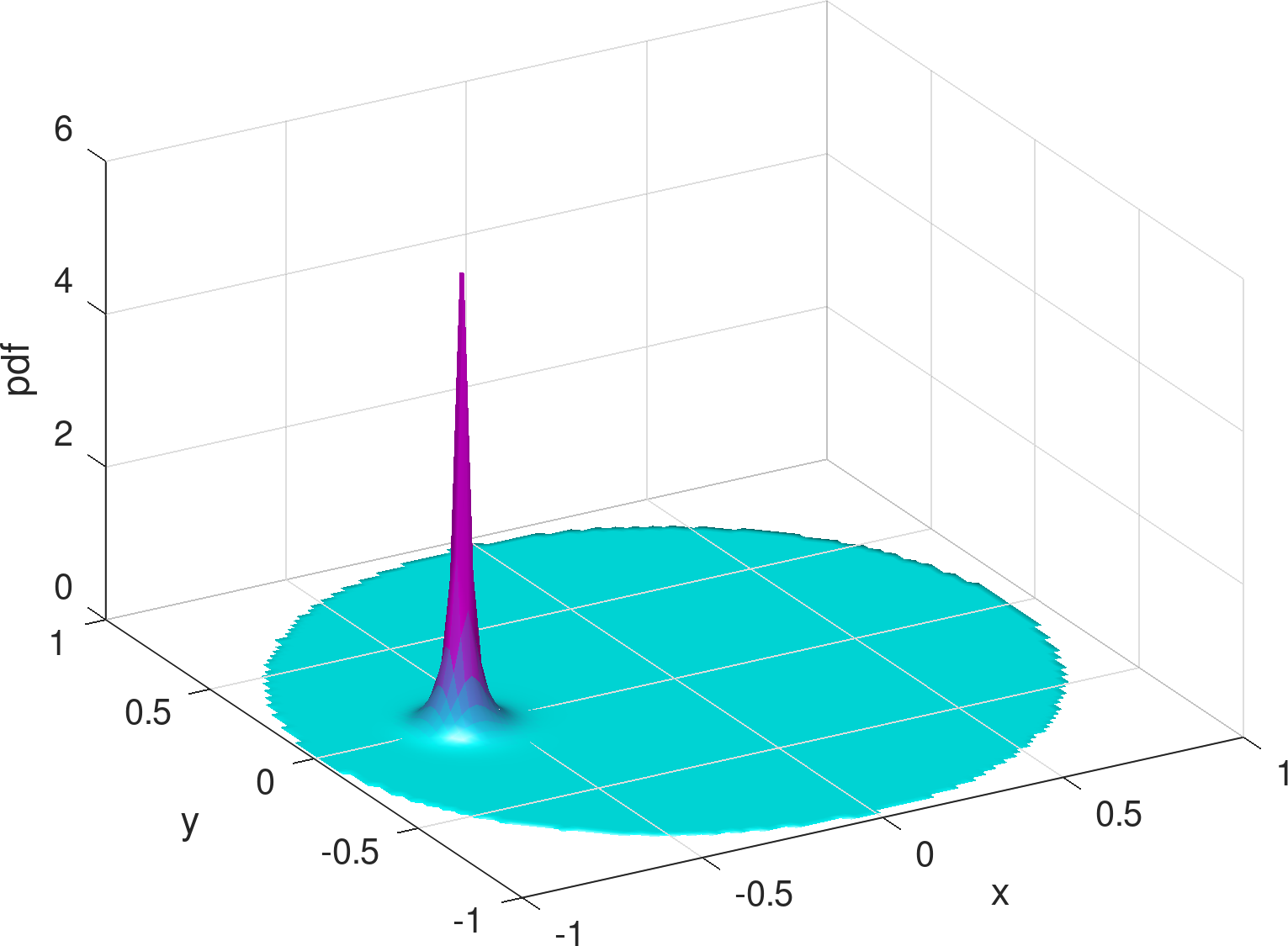}}
 \subfigure[]{
 \label{fig:g_m:i}
 \includegraphics[width=5.4cm]{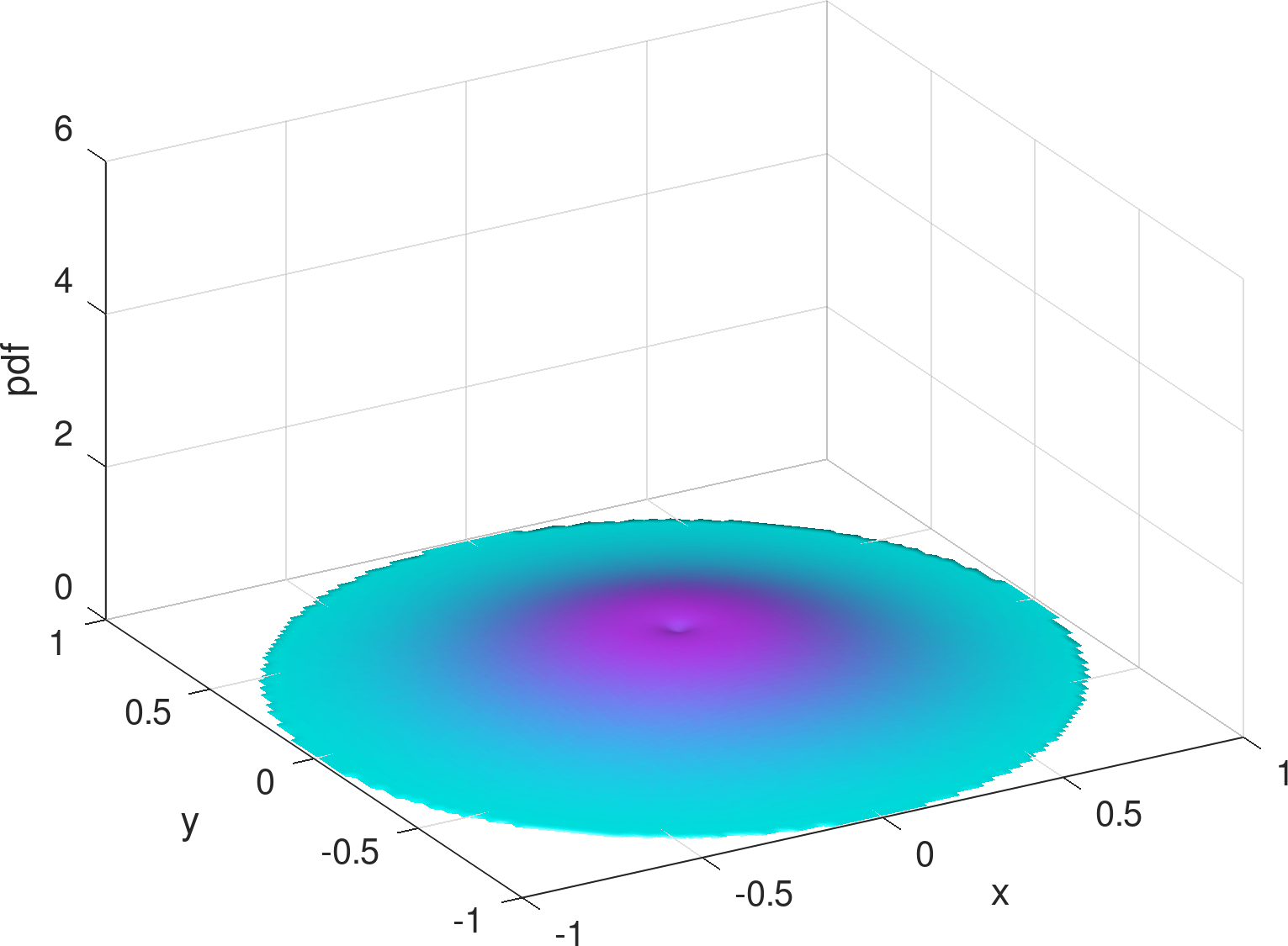}}
 \subfigure[]{
 \label{fig:g_m:j}
 \includegraphics[width=5.4cm]{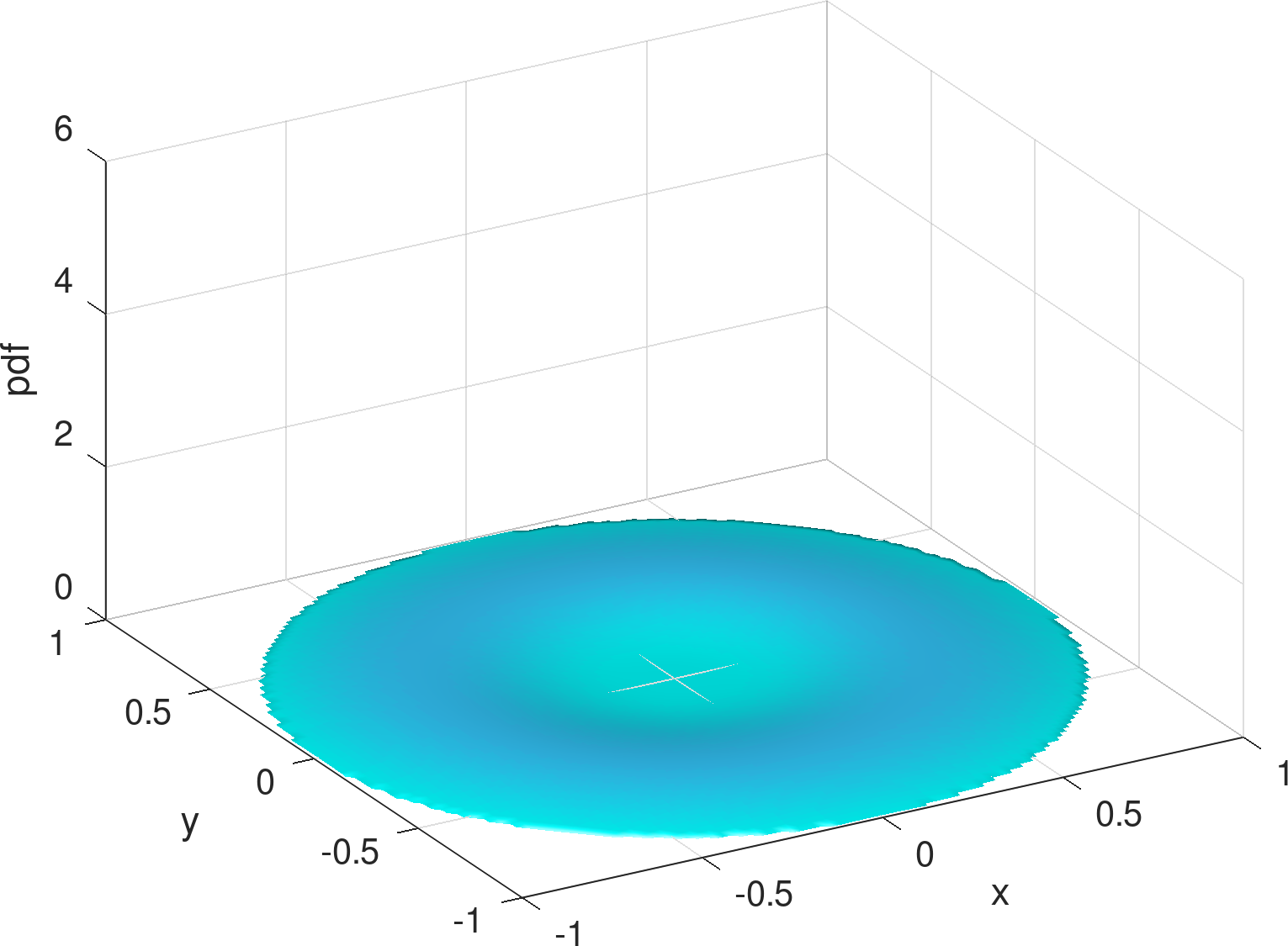}}
 \subfigure[]{
 \label{fig:g_m:k}
 \includegraphics[width=5.4cm]{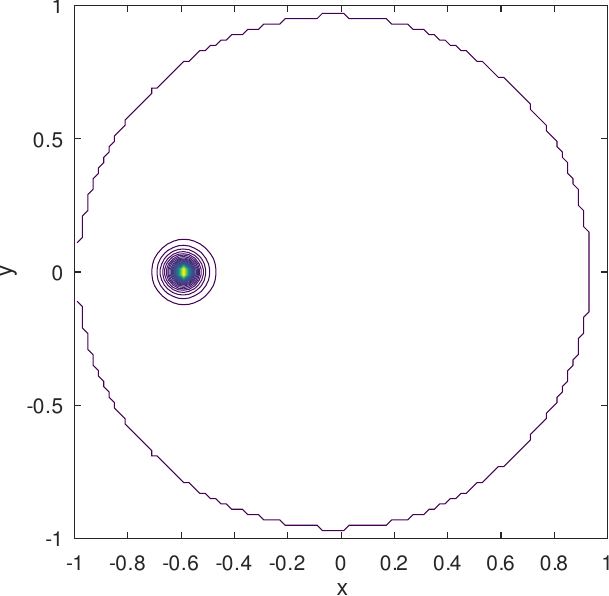}}
 \subfigure[]{
 \label{fig:g_m:l}
 \includegraphics[width=5.4cm]{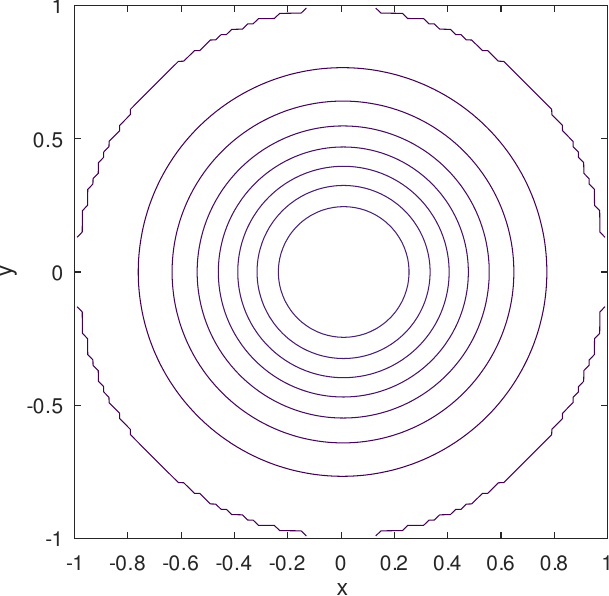}}
 \subfigure[]{
 \label{fig:g_m:m}
 \includegraphics[width=5.4cm]{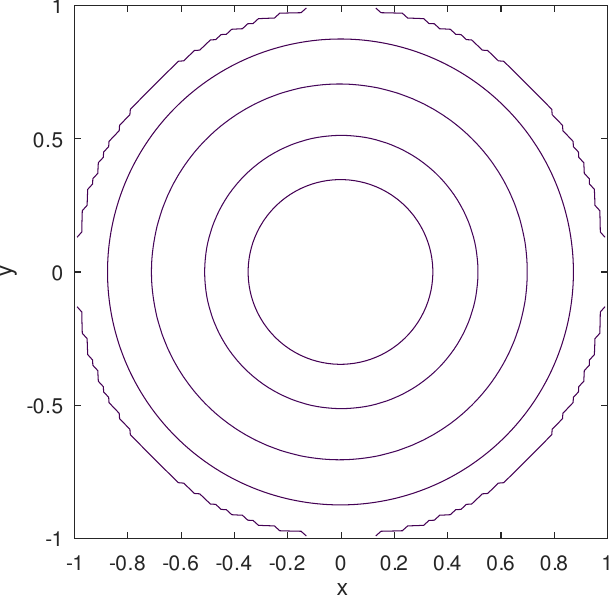}}
 \caption{Constituent distributions of the Gaussian and M\"obius mixture models: (1) top two rows corresponding to Figures \ref{fig:gau_bvm_mob}(a)(d); (2) bottom two rows corresponding to Figures \ref{fig:gau_bvm_mob}(c)(f).}
 \label{fig:g_m}
\end{figure*}

The inability of the Gaussian mixture to accurately mimic the target distribution can be understood by considering the shape of its constituent distributions. As an example, the three Gaussian distributions (i.e., $K=3$) plotted in the first two rows of Figure \ref{fig:g_m} are superposed to produce the result shown in Figure \ref{fig:gau_bvm_mob}(a)(d). While the peak at the charger's location is not lost and can be reproduced (partially) by one Gaussian component, the shape fitted is subpar with a height lower than supposed. The remaining background dome of the distribution is then insufficiently reconstructed by the two remaining Gaussians (Figures \ref{fig:g_m}(b)(c)(e)(f)), which are placed in a haphazardly asymmetric manner. The shape induced by the Gaussians lacks the requisite symmetries as in Figure \ref{fig:exact}, leaving a lot to be desired. %

On the other hand, the three M\"obius distributions, which correspond to the bottom two rows of Figure \ref{fig:g_m} and combine to produce Figure \ref{fig:gau_bvm_mob}(c)(f), are able to depict the target distribution in an interpretable, component-wise manner, like the Gaussian mixture, but with considerably more faithful matching of the shape. One M\"obius distribution is sufficient to cleanly capture the charger peak (Figures \ref{fig:g_m}(g)(j)) and the remaining two for the main RWP dome $f_{\mathrm{RWP}}(z)$ (Figures \ref{fig:g_m}(h)(i)(k)(l)). Unlike the Gaussians, these two components remain supported on the disk, retaining the rotational symmetry of $f_{\mathrm{RWP}}(z)$ and avoiding the stretched surface of the mixed Gaussians subject to bell-shaped convexity.\footnote{Mixtures of truncated Gaussians, although retaining the bounded nature of the disk support, exhibit similar asymmetric domes to the Gaussians without truncation.} Crucially, this interpretability of MDNs is relinquished when NFs are used.

\begin{figure*}[!t]
 \centering
 \subfigure[]{
 \label{fig:nf:a}
 \includegraphics[width=5.6cm]{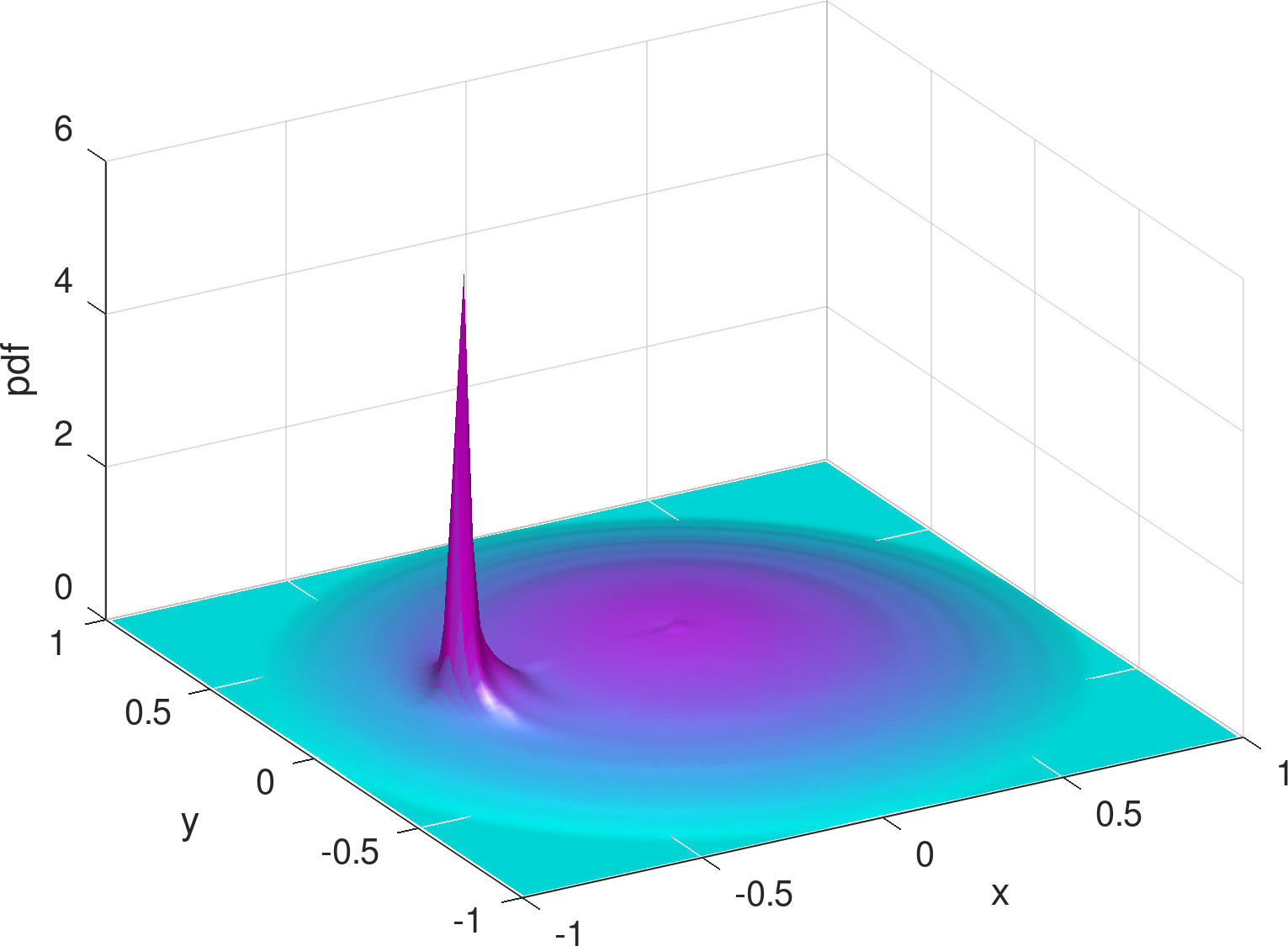}}
 \subfigure[]{
 \label{fig:nf:b}
 \includegraphics[width=5.6cm]{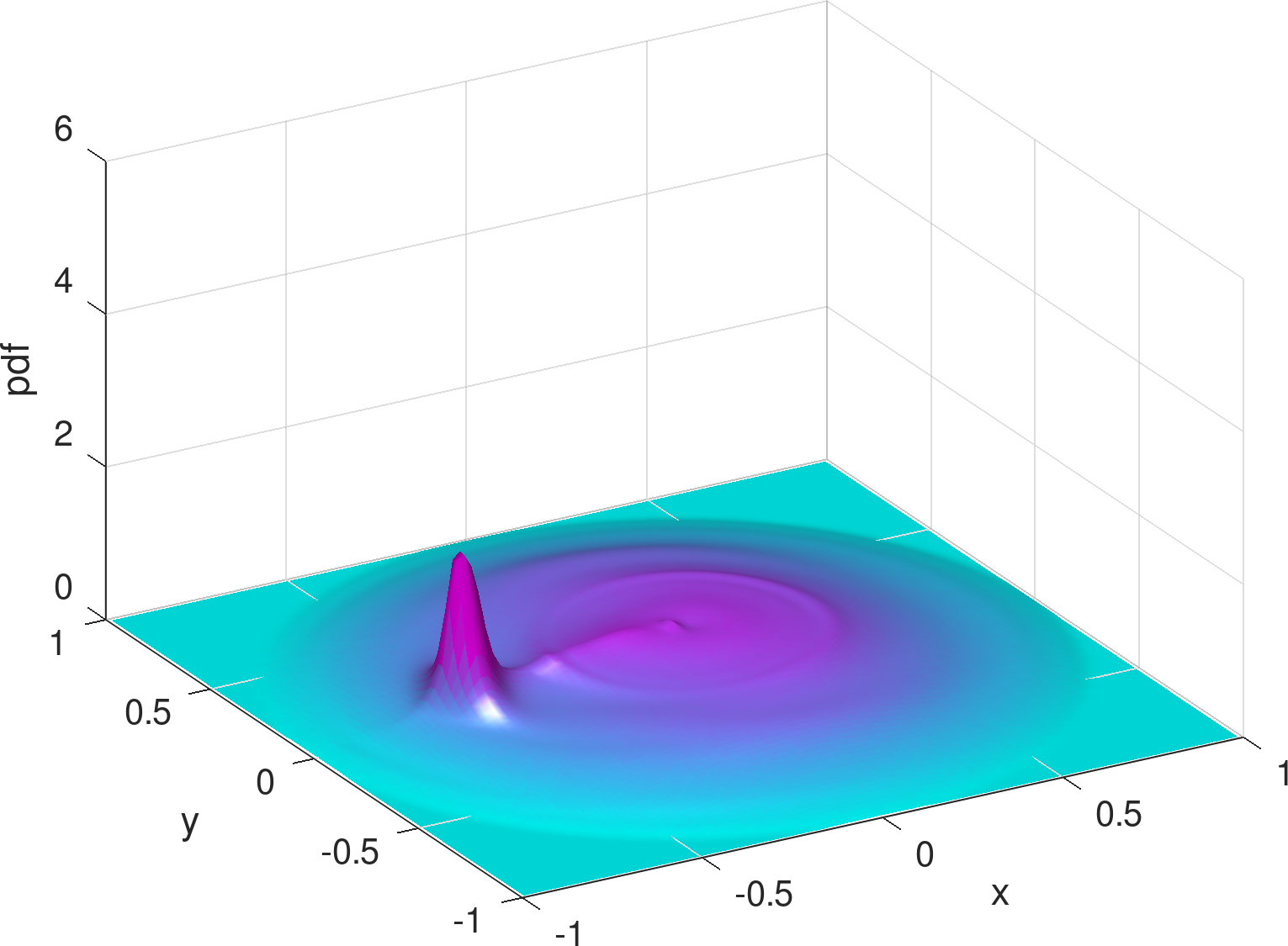}}
 \subfigure[]{
 \label{fig:nf:c}
 \includegraphics[width=5.6cm]{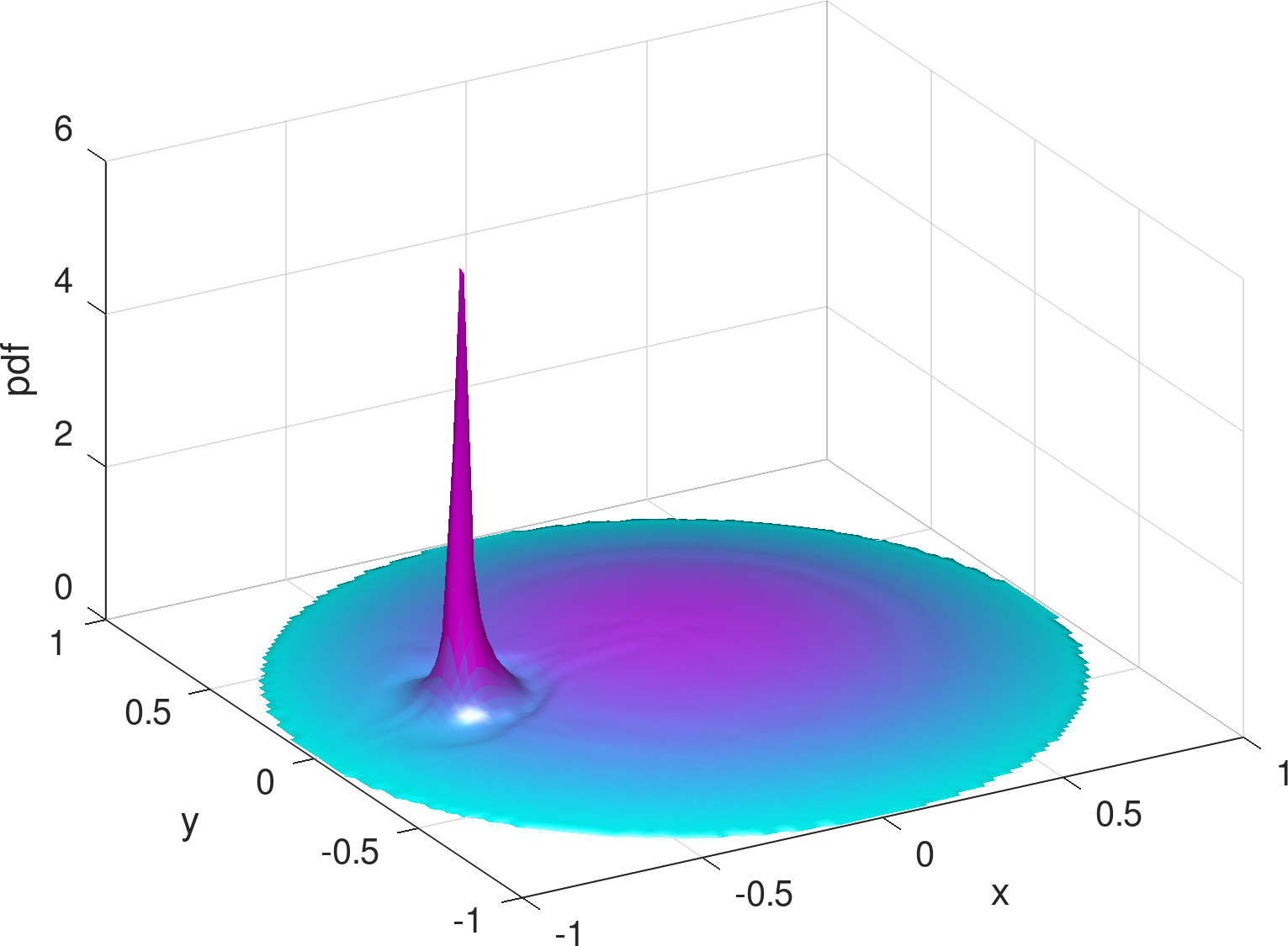}}
 \subfigure[]{
 \label{fig:nf:d}
 \includegraphics[width=5.6cm]{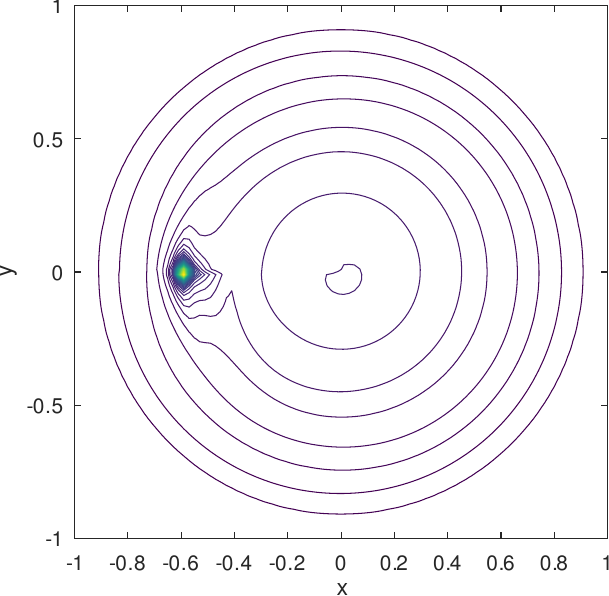}}
 \subfigure[]{
 \label{fig:nf:e}
 \includegraphics[width=5.6cm]{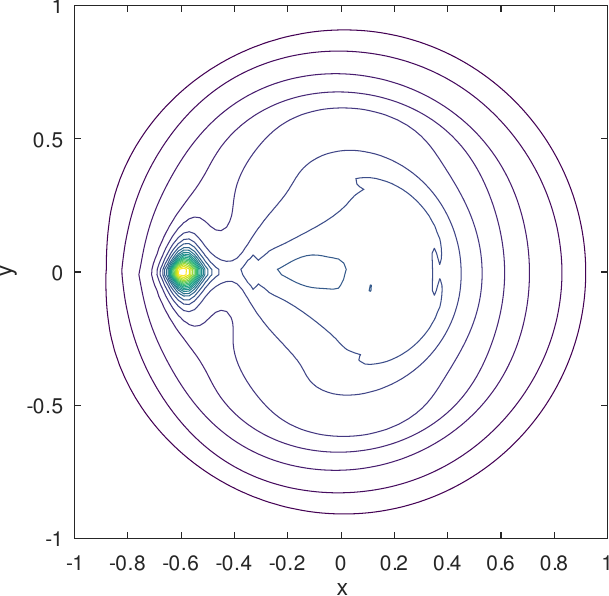}}
 \subfigure[]{
 \label{fig:nf:f}
 \includegraphics[width=5.6cm]{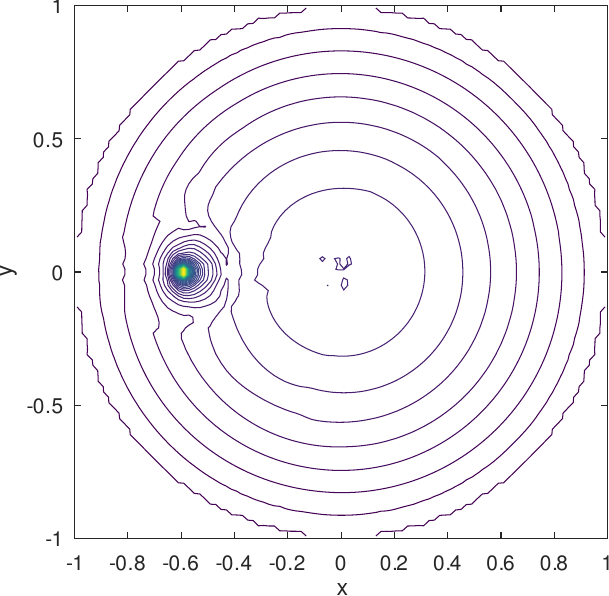}}
 \caption{NFs with four coupling layers (i.e., $L$=4), plotted from a training set of 24 points: (a)(d) A good fit from RNVP without truncation, with \(\text{MSE}=0.00098\) and \(\text{GKL}=0.00185\); (b)(e) A bad fit from RNVP without truncation, with \(\text{MSE}=0.00479\) and \(\text{GKL}=0.00470\); (c)(f) NSF, with \(\text{MSE}=0.00003\) and \(\text{GKL}=0.00009\).}
 \label{fig:nf}
\end{figure*}

Figure \ref{fig:nf} shows that the untruncated RNVP model (representative of both RNVP variants) can indeed recover the broad target shape, although its fit varies unstably across runs (Figures \ref{fig:nf}(a)(d) versus \ref{fig:nf}(b)(e)), whereas NSF reproduces even the finer details (e.g., dips around the charger) with near-perfect accuracy. Nonetheless, both the RNVP and NSF models act as a monolithic ``black box'' of transformation, providing no attributable components to tease apart the charger's attraction effect from the background RWP mobility. The M\"obius MDN, by contrast, achieves accuracy still comparable with the NFs while retaining both interpretable decompositions and parameter efficiency. Note that the strong NSF fit shown here is not always guaranteed; as discussed next, when there are fewer (e.g., five) training points, which is not uncommon in practice due to costly simulations, NSF can induce poor predictions.

\subsection{Prediction under Data Scarcity}\label{sec:prediction}

To examine a more data-scarce regime, we train the most competitive model of each type, i.e., M\"obius MDN and NSF, on an even smaller training set of five points only, with $d=0.2$ and $\theta_c=\pi$ constantly while $(d', r_c) \in \{(0.1, 0), (0.2, 0), (0.2, 0.2), \allowbreak (0.2, 0.6), \allowbreak (0.2, \allowbreak 0.8)\}$. We then let the two (sufficiently trained) models predict a held-out setting of $(d',r_c)=(0.2,0.5)$, which is an interpolation task.

\begin{figure*}[!t]
 \centering
 \subfigure[]{
 \label{fig:mob_nsf:a}
 \includegraphics[width=5.6cm]{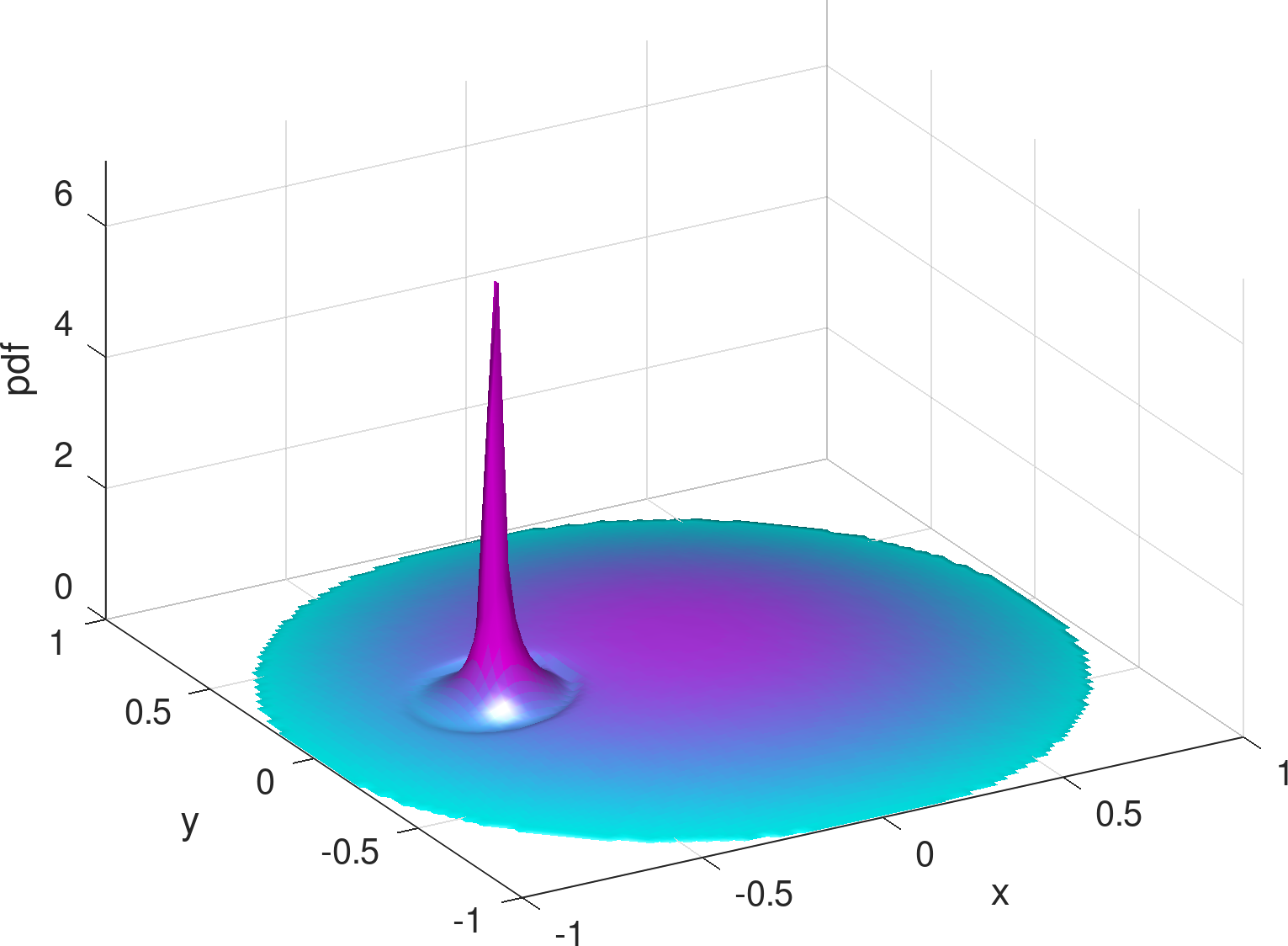}}
 \subfigure[]{
 \label{fig:mob_nsf:b}
 \includegraphics[width=5.6cm]{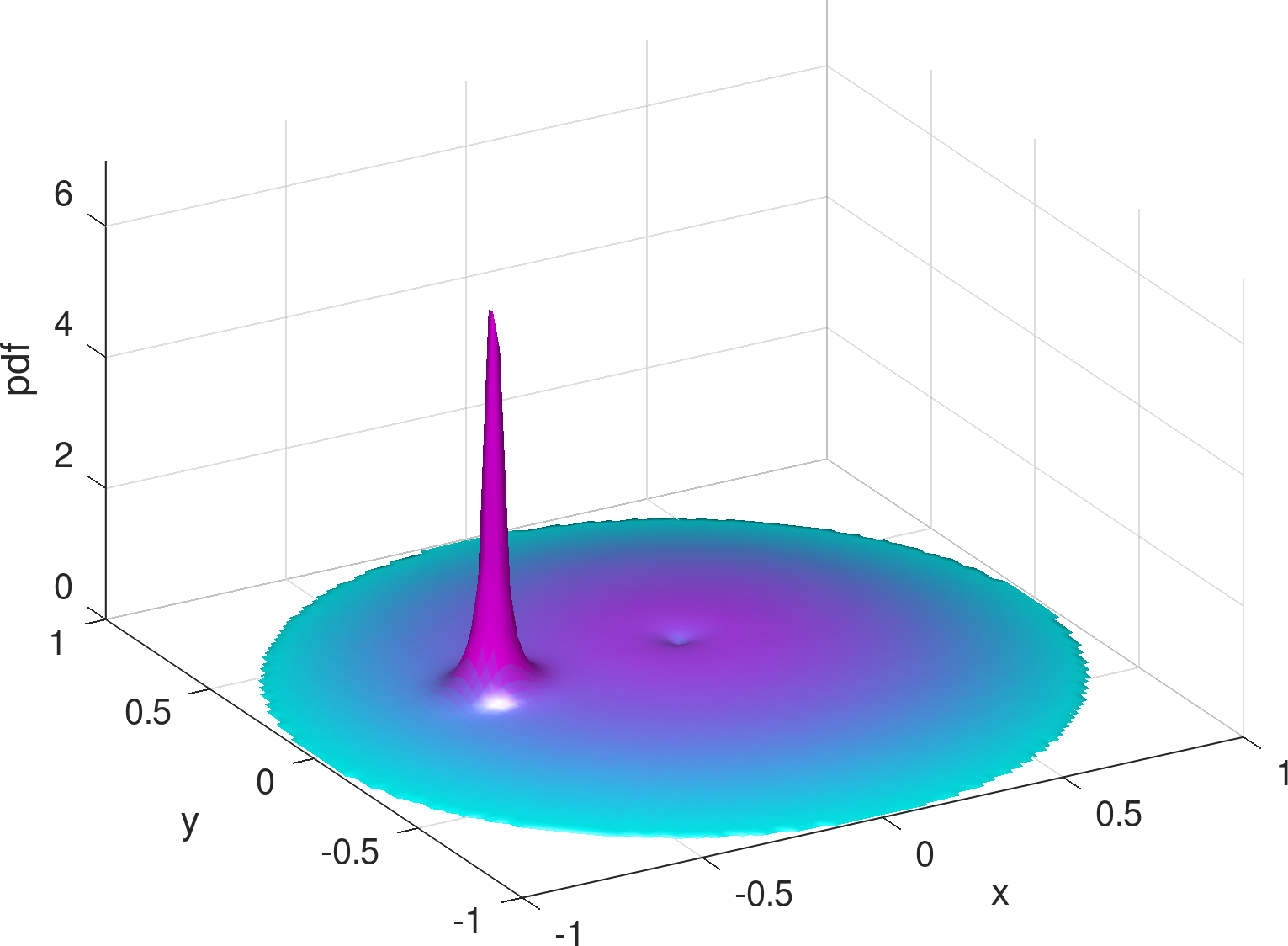}}
 \subfigure[]{
 \label{fig:mob_nsf:c}
 \includegraphics[width=5.6cm]{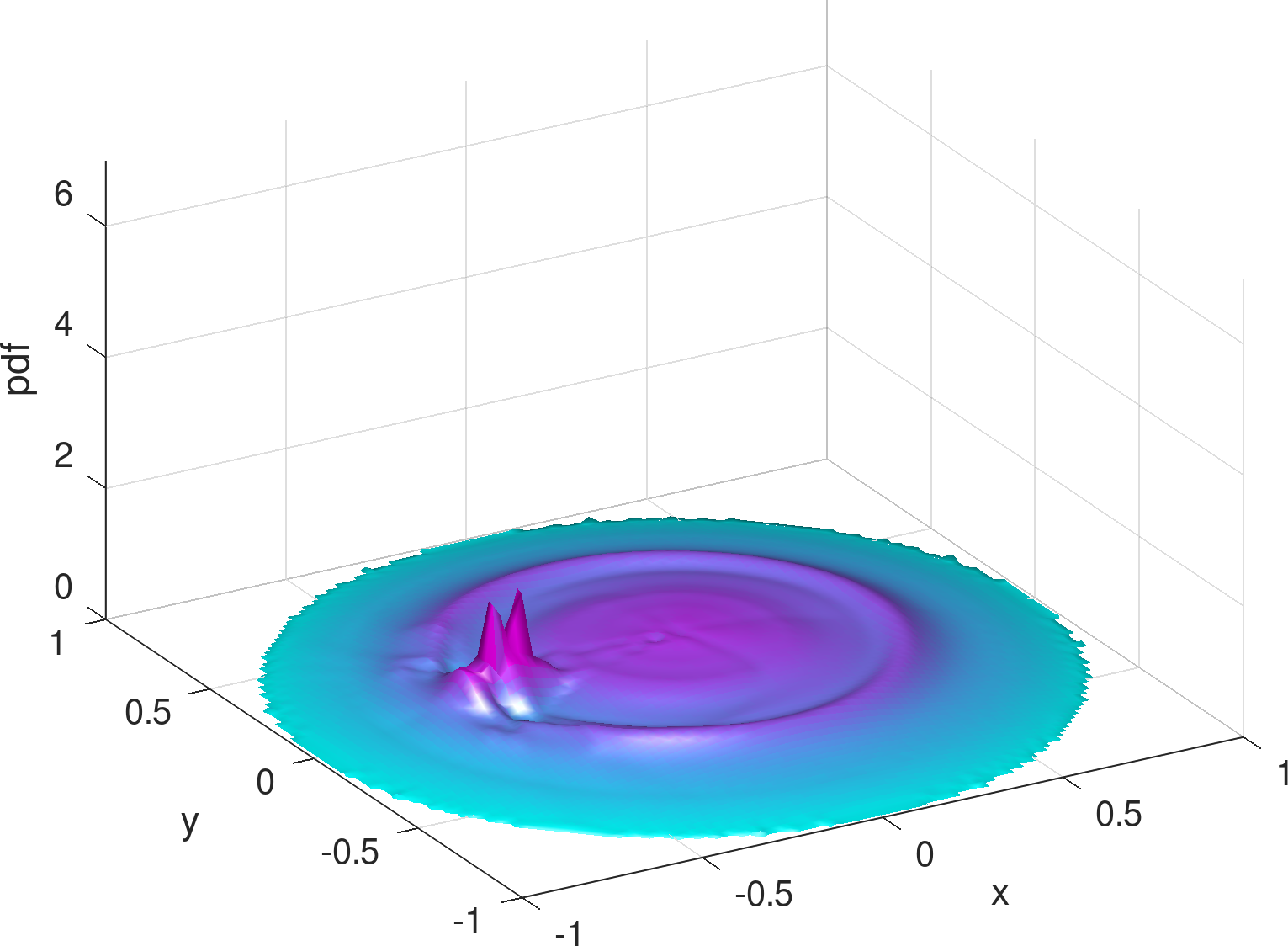}}
 \subfigure[]{
 \label{fig:mob_nsf:d}
 \includegraphics[width=5.6cm]{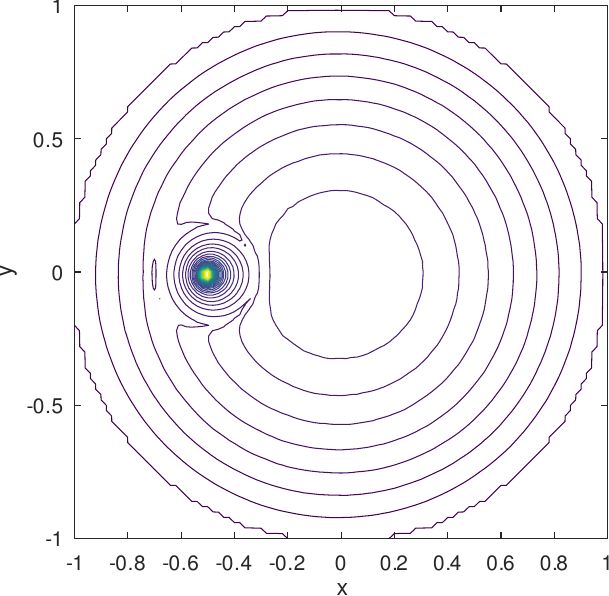}}
 \subfigure[]{
 \label{fig:mob_nsf:e}
 \includegraphics[width=5.6cm]{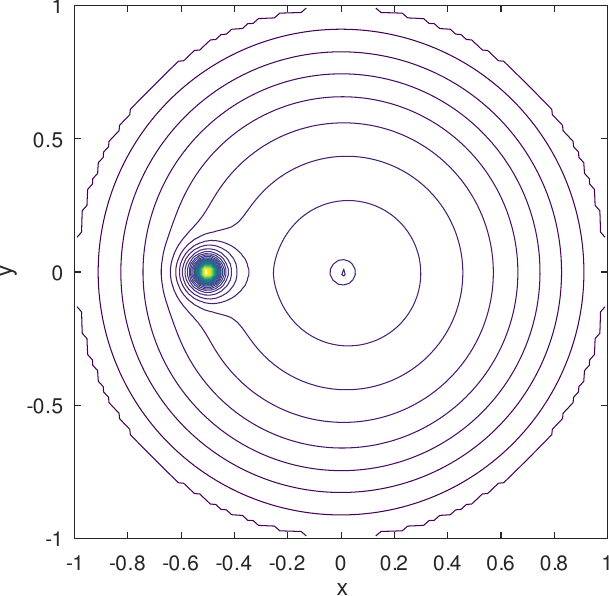}}
 \subfigure[]{
 \label{fig:mob_nsf:f}
 \includegraphics[width=5.6cm]{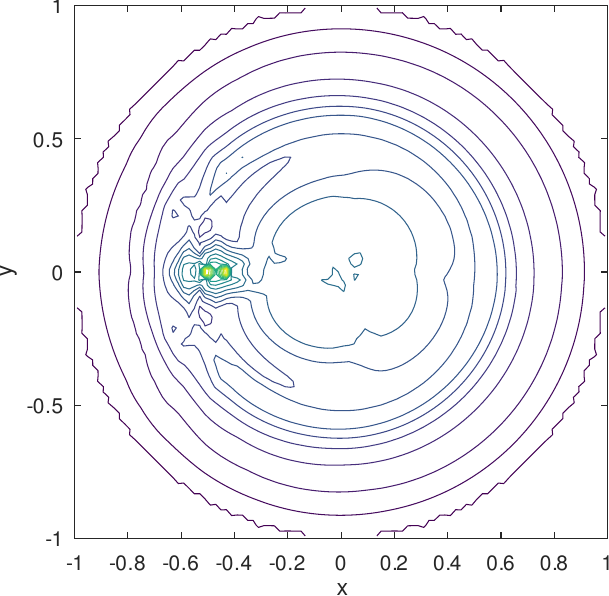}}
 \caption{M\"obius MDN with three components (i.e., $K$=3) versus NSF with four coupling layers (i.e., $L$=4), both trained with 5 points only and used for predicting the held-out point with $(d',r_c)=(0.2,0.5)$: (a)(d) Exact distribution; (b)(e) M\"obius MDN, with \(\text{MSE}=0.00192\) and \(\text{GKL}=0.00209\); (c)(f) NSF, with \(\text{MSE}=0.01314\) and \(\text{GKL}=0.01569\).}
 \label{fig:mob_nsf}
\end{figure*}

Figure \ref{fig:mob_nsf} provides an illustrative example of NSF's vulnerability to erratic spatial predictions in this sparse-data scenario. That is, the prediction from NSF (assuming $L=4$) becomes marred by spurious spikes and concentric oscillatory artifacts (Figures \ref{fig:mob_nsf}(c)(f)), as also reflected in its higher MSE and GKL values. In constrast, the M\"obius MDN (with $K=3$) preserves the smooth background dome and the localized charger effect (Figures \ref{fig:mob_nsf}(b)(e)), producing a closer match to the exact distribution (Figures \ref{fig:mob_nsf}(a)(d)), with \(\text{MSE}=0.00192\) and \(\text{GKL}=0.00209\). The results show that the structural bias of the M\"obius kernels enables a more plausible prediction from sparse observations, whereas the more expressive, parameter-intensive, and relatively data-intensive NSF can be sensitive to a limited coverage of the conditioning space, requiring additional simulation runs to recover comparable accuracy.

\subsection{Model Training and Its Dynamics}\label{sec:training}

Each model is trained for 20,000 epochs to guarantee convergence across all configurations and each epoch includes (only) one batch. Quantitative metrics (e.g., Table \ref{tab:g_vs_m}, Figures \ref{fig:train_perf_mdn} and \ref{fig:pred_perf_mdn}) are recorded at the epoch when the training loss is minimized. For the optimizer, we use Adam (adaptive moment estimation) instead of SGD (stochastic gradient descent), as we observed the latter (even with tuned momentums) would be prone to underfitting during training \cite{wilson2017marginal}. In addition, the learning rate is set to 0.0001 for both MDNs and NFs, which is empirically optimal for the models.%

\begin{figure*}[!t]
 \centering
 \includegraphics[width=\textwidth]{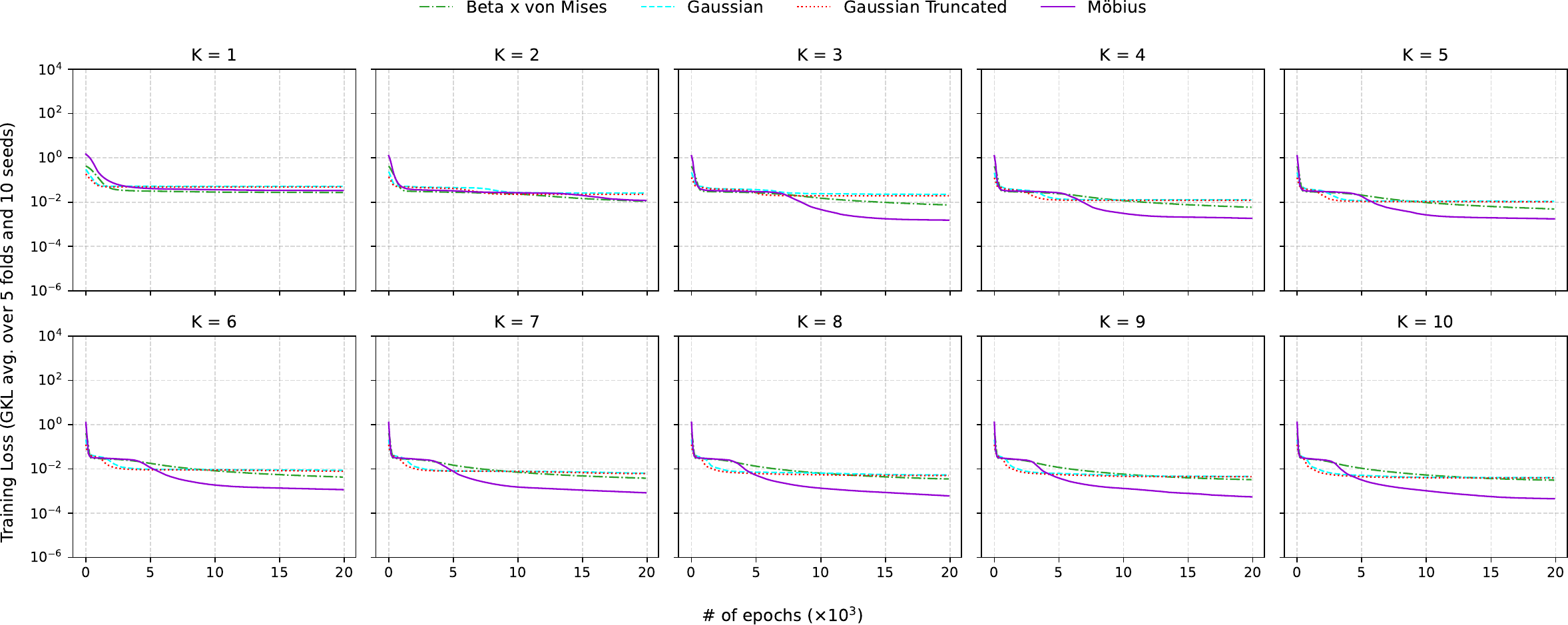}
 \caption{Evolution of the training loss (GKL) for the MDN models across varying numbers of mixture components ($K$). Each curve depicts the average loss over 5 folds and 10 independent runs.}
 \label{fig:train_loss_mdn}

 \vspace{2em}

 \includegraphics[width=\textwidth]{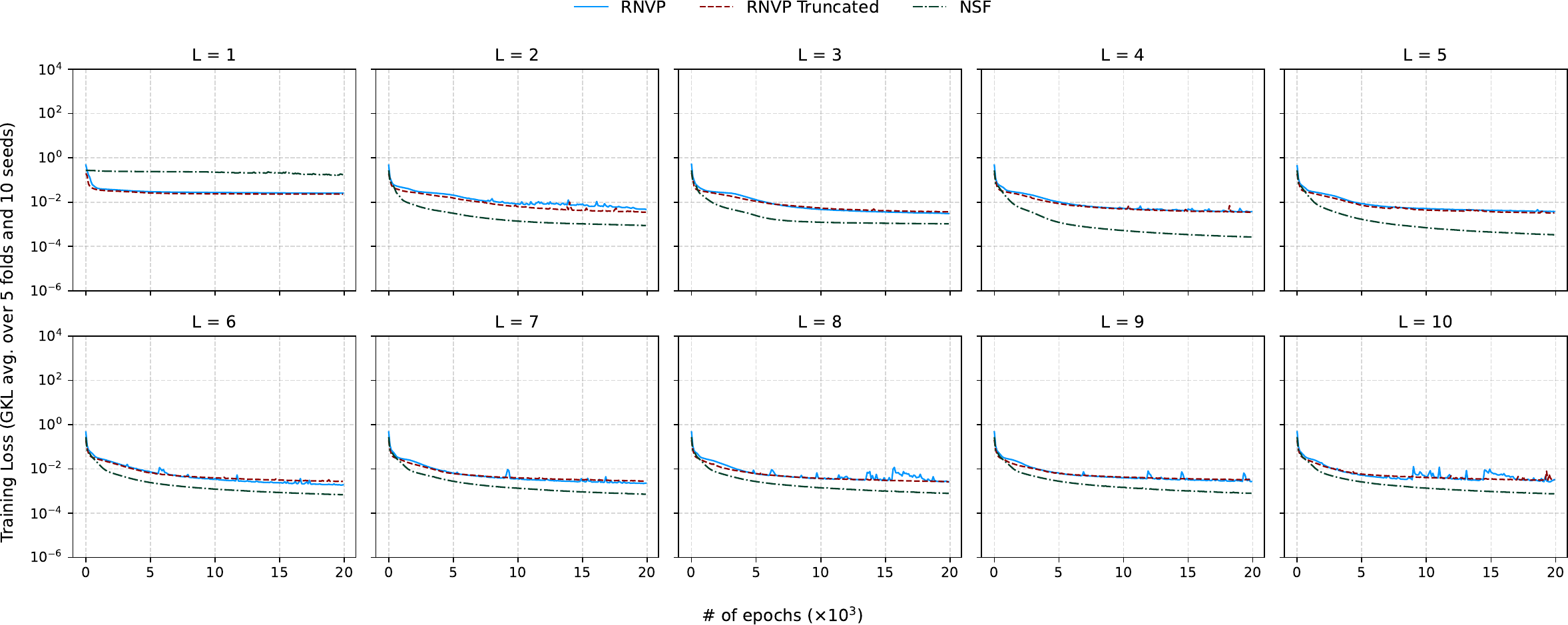}
 \caption{Evolution of the training loss (GKL) for the NF models across varying numbers of coupling layers ($L$), with curves averaged over 5 folds and 10 runs.}
 \label{fig:train_loss_nf}
\end{figure*}

The training dynamics are further illustrated by the loss evolution curves in Figures \ref{fig:train_loss_mdn} and \ref{fig:train_loss_nf}, which trace the training loss (GKL divergence) averaged over 5 folds and 10 independent runs. As seen in Figure \ref{fig:train_loss_mdn}, the M\"obius mixture consistently descends to a lower floor than the Gaussian models. Its curves often level off for a time before a second descent; this pause is shorter at larger $K$, suggesting an initial fit to the RWP dome followed by a component allocated to the charger peak, which can be expedited with extra components. While dropping not so rapidly as the Gaussians in early epochs, the beta-von-Mises model then continues a slow descent rather than flattening as the Gaussians, yet remains at a higher level than the M\"obius model due to its structural limitations. Regarding the NFs, Figure \ref{fig:train_loss_nf} shows that NSF, once $L\geq 2$, follows a smooth descent to a markedly lower floor than either RNVP variant. Both RNVP models improve from $L=1$ but then occupy a similar, higher plateau whether truncated or not. At certain larger values of $L$, the untruncated RNVP curves exhibit isolated spikes, indicating occasional training instability.

\subsection{Inference Computation Performance}\label{sec:inference}

The benefits of a trained model can be better appreciated in the context of the running time required to determine the distribution for a given set of $\lambda=(d, d', r_c, \theta_c)$. We use an Intel i7-8700 system with Windows 11 and Windows Subsystem for Linux (WSL) 2 hosting Ubuntu 24.04.3. No GPU is used, and no explicit parallelism is programmed or requested, hence the running times quoted here are for single-core sequential execution:
\begin{itemize}
    \item a single numerical evaluation of Equation \ref{eq:f_z} in MATLAB R2024b (running natively on Windows 11 without any MATLAB parallelism directives; defaults used for the numerical evaluation accuracy, e.g.. those impacting integration), takes approximately \textbf{108 minutes},
    \item a single simulation of the system behavior using a custom-built C++ simulator for a total traveled distance of $50\times 10^6$ units where the unit of distance is the disk radius (\texttt{gcc 13.3.0},  \texttt{-O3} optimization, \texttt{boost 1.83} library, on WSL2) takes approximately \textbf{11.5 minutes}, 
    \item a single inference/prediction invocation with the MDN model (\texttt{python 3.12.3}, \texttt{numpy 2.4.4}, \texttt{torch 2.12.0} \texttt{+cpu}, on WSL2, without any GPU support), takes approximately \textbf{7 milliseconds} with coefficient of variation of 0.14, 
    \item a single inference/prediction invocation with the RNVP NF model takes approximately (for 1 to 10 layers, respectively) \textbf{17, 21, 23, 28, 31, 34, 35, 38, 42, and 46 milliseconds} with coefficients of variation of 0.16 or less, and, 
    \item a single inference/prediction invocation with the NSF NF model takes approximately (for 1 to 10 layers, respectively) \textbf{14, 18, 22, 28, 32, 37, 46, 48, 53, and 60 milliseconds} with coefficients of variation of 0.14 or less. 
\end{itemize}

Clearly the produced models are orders of magnitude faster than the simulation and numerical computation options. The Python inference times do not include the model loading time, which is assumed already present in memory. In any application the cost of loading the model is amortized across the multiple invocations to the inference function. It is worth noting that, while the simulation time can be reduced proportionally to reduce the simulated distance traveled (currently, $50\times 10^6$ units), it was observed that the measured density deteriorates significantly if the density measurement intervals were reduced below $10^6$ units.

\section{Related Work}\label{sec:related}

Broadly speaking, our objective is the approximation of spatial density distributions that arises in the study of mobile networks. We use a particular non-trivial density as our benchmark, that has been derived from file-grained mobility under charger and energy constraints. The literature of mobility modeling restricts the domain of the distributions up to a circular boundary. As such, the problem under consideration falls under the general category of {\it inverse problems}, i.e., problems where the observed data are assumed to come from a function whose parameters need to be derived. Solving inverse problems using learning algorithms has met with some success in various domains in the natural sciences \cite{xu2021solving,cuomo2022scientific,gilmer2017neural}. The shape of the density function we are interested in as examples have been encountered in \cite{9678073} and \cite{8653318} and suggest a rather well-behaving, continuous and smooth function with the notable exception of the singular point at the exact location of the charger(s). Nevertheless, it is difficult to find a tractable density function that fits the shape while being analytic. The derivation of a closed form is considered as valuable, not only for accelerating the simulation process but also for facilitating further mathematical analyses of network performance (e.g., wireless capacity \cite{8653318}). Hence, we disregard the class of sampling-based generative models (e.g., \cite{gammelli2022recurrent}) that can only numerically output the density value given the location.

The source of complications is the existence of a boundary. While the boundary is meaningful for mobility studies, as it approximates the boundary of e.g., an urban environment, a neighborhood, etc., this turns out to introduce a challenge when approximating the distribution via learning techniques. Learning techniques are used to provide simple, easy-to-compute as needed distribution models. The evaluation of the density at any point of a circularly bounded domain should be fast to execute, if such density distributions have any hope to be used as subroutines in optimization, e.g., resource placement, problems. Hence, our quest for alternative models is motivated by our intent to produce computationally ``light'' generalizable means of deriving density estimates. The question is how to derive such models from a small set of simulated data, since even the simulation is computationally expensive. 

We begin with the classical and analytically convenient class of mixture models. %
To our understanding, spatial density learning using finite-supported, non-Gaussian mixtures for mobility modeling was, until recently, still lacking in the mobile networking community. 
Our recent work has addressed convex-polygonal (e.g., triangular and quadrilateral) and square supports with kernels other than the M\"obius distribution \cite{feng2025irregular,zhao2025biased}. Those geometries and kernels are distinct from, and complementary to, the disk-supported M\"obius mixtures developed here.

There are related studies on Gaussian-based mixture density learning for urban transportation, whose interests or assumptions are nevertheless at variance with our work. For instance, in order to alleviate traffic congestion in urban areas, the GPS traces of over 1,000 taxis in a city are analyzed in \cite{tang2019estimating}, where a Gaussian mixture model is adopted to model the taxis' location distribution for hotspot evaluation. The mobility considered there is constrained, in the sense that the traces are limited to graph-like road networks. This is different from our assumption of a bounded circular movement area in the Euclidean plane. 

In \cite{cho2011friendship}, the patterns of cellphone users, who have transitional states of location driven by temporal periodicity and social interplay, are modeled by mixing pairs of Gaussians (accounting for periodicity), together with a power-law distribution (for social dynamics). In their example, the spatial model assumes a square-shaped support for mobility, but it is unclear how the potential density leakage beyond the boundary is addressed. Also, the number of component distributions is supposedly fixed in \cite{cho2011friendship}. In \cite{lichman2014modeling}, Gaussian kernels with adaptive bandwidths (parameterizing the correlation matrices) are introduced for estimating spatial density of mobile users at different scales. The bandwidths are adapted by the kernel density estimation method to account for density discrepancy between highly (e.g., roads) and sparsely (e.g., deserts) populated areas. Like \cite{tang2019estimating,cho2011friendship}, the focus is mainly on accommodating the heterogeneous dependence of density on locations, with no concern or solution for fitting specifically confined movement areas.

It is worth mentioning another type of mixture density learning problem that is often considered in autonomous driving and robot navigation, e.g., \cite{makansi2019overcoming, zhou2023query, choi2018uncertainty}, which appear to be related but are in essence distinct from our scenario. There, the basic idea is to use mostly Gaussian and occasionally non-Gaussian (e.g., Laplacian \cite{zhou2023query}) mixture models to predict near-future trajectories of vehicles or robots in the locality (e.g., at an intersection) given a relatively short past history (e.g., within seconds), such that the moving agents can navigate efficiently and safely without collision. As can be seen, the main concern of such studies is prescriptive route planning or maneuver control at relatively small scales in both spatial and temporal aspects, or equivalently, involving one or at most a few pairs of random ``waypoints'' in the whole learning process. In contrast, the (disk-shaped) movement area we have assumed can be as broad as a city, and the span of a nodal trajectory simulated corresponds to the node moving for extended periods of time such that a steady state can be reached. Applying  short-term prediction methods is different from our steady-state based problem formulation.

The application of non-Gaussian mixtures, with finite or semi-finite supports, appears to be more common in learning non-geographic density distributions. Such applications have occurred in various domains like astrophysics \cite{wang2022likelihood}, economics \cite{delong2021gamma}, and power engineering \cite{zhang2020improved}. For instance, beta distributions (with a support of $[0, 1]$) are mixed in \cite{wang2022likelihood} to infer cosmological parameters of one-dimensional CDM (cold dark matter) models. The betas are also employed in \cite{zhang2020improved} for an improved MDN such that the univariate wind power output of wind farms can be forecast with no issue of density leakage. In \cite{delong2021gamma}, an MDN using the multivariate Gamma distribution (having a support of $[0, +\infty)$ for each marginal variable) as its kernel is proposed to model the distribution of insurance claim amounts. The methods mentioned above are certainly not applicable to our scenario of mobility modeling for mobile networks.

We encounter challenges because the choice of kernel turns out to be non-trivial especially for densities that are radially symmetric and have support strictly inside a Euclidean disk. Naturally, we consider truncated distributions, but we look into more direct approaches, such as M\"obius-based distribution mixture learning models, inspired by the work of \cite{sym11081030}, where a new class of disk-supported distributions (including the beta type \Rn{3} M\"obius distribution) is proposed to model the bivariate density of wind speed and wind direction over Marion Island. Their work focuses on deriving fitted distribution forms for accurate wind modeling, making no mention of density mixture for potential fitness improvement. Inspired by \cite{sym11081030}, in this paper we extend to mixtures of M\"obius distributions. A remarkable outcome of our extension is that the results become more ``interpretable'' in the sense that the M\"obius distributions reveal local and global symmetries, something not possible with Gaussians (truncated or not). For example, our results show one of the M\"obius distributions of the mixture capturing the presence of a RWP component (Figure \ref{fig:g_m} and related text) which is also present as a component ($f_{\mathrm{RWP}}(z)$) in the analytical form (Equation \ref{eq:f_z}) of the distribution!  

A natural next step is to search for expressive density estimators without necessarily relying to mixtures. Normalizing flows are a notable means to accomplish this objective. In principle a flow could map a simple base measure onto the observed mobility density. The geometry of the domain, however, imposes severe constraints:

Flows designed for closed manifolds (Möbius transformations on spheres or tori \cite{rezende2020normalizing}) act on spaces without boundary; they cannot be transplanted to a bounded Euclidean disk. Hyperbolic flows on the Poincaré disk \cite{bose2020latent} place the boundary at a conceptual infinity, whereas mobile nodes in the RWP (and similar) models can reach the Euclidean boundary in finite distance/time.
Mapping an unbounded Gaussian onto a bounded disk requires a squashing function (e.g., $  r\mapsto\tanh(r)  $). Because the mobility density we are trying to approximate places positive mass near the boundary, the inverse map sends those points to infinity, producing vanishing gradients, infinite Jacobian determinants and numerical instability. We witnessed this behavior in preliminary experiments.

A “true” flow on the disk would therefore need a base distribution already supported on the disk together with bounded bijections (e.g., radial neural splines on $[0,1]$ and circular splines on $[-\pi,\pi]$). Constructing and training such a model from scratch constitutes a substantial engineering effort. Thus the modeling path we follow is deliberate: mixture models supply the closed-form density required for analysis; normalizing flows, while more flexible in unconstrained domains, confront geometric and numerical obstacles that currently outweigh their potential gains for this particular inverse problem. As a potentially powerful tool, especially for expressing ``steep'' transitions at the boundaries, we consider NSF \cite{durkan2019neural} in addition to ``standard'' RNVP \cite{dinh2017density} normalizing flows, as NSF can, at the cost of more computation, express a broader set of monotonic functions compared to the affine maps of RNVP.

\section{Conclusions}\label{sec:conclusions}

Assume you are in the position that, after having completed a few simulations, you recognize that the results exhibit a pattern such as the one in Figure \ref{fig:exact}. Using machine learning tools of your choice to predict the results for other parameter settings, you are given a model which decomposes the distribution as in Figure \ref{fig:g_m}(g)-(l)
and one model that produces Figure \ref{fig:mob_nsf}. It is unlikely that you would pick the latter if your intention was to understand  the factors that come into play in the analytical description of the distribution---assuming you are after a description similar to Equation \ref{eq:f_z}. Simply put, interpretability, allows humans to develop insights and further analytical insights. 

Our work reveals that the M\"obius distribution, through its  ability to capture symmetries, having finite support ``baked in'' (rather than haphazardly enforced via truncation methods), provides an interpretable, parsimonious, and accurate tool to learn steady state distributions of mobile nodes on a disk. Normalizing flows are serious contenders but lacking inerpretability and suffering when training data are scarce. The benefits of developing M\"obius-based mixture models can be capitalized on when an overarching optimization goal, e.g., the placement of points of interests such as chargers, requires the quick calculation of entire density distributions.

A natural next step is to replace some of the density-influencing parameters with stochastic counterparts. As an example, the distance $d$ which acts as a proxy of energy availability upon departure from a waypoint, can be replaced by distributions with a mean $d$, and there may be benefit to also include higher moments of such distributions in the modeling process.

\bibliographystyle{elsarticle-num}
\def\urlprefix{}
\renewcommand{\url}[1]{}
\let\newline\relax
\hypersetup{hidelinks}
\bibliography{refs,zotero}

\end{document}